\documentclass{article}
\input epsf
\begin{document}

\title{\textbf{Transition Matrix Monte Carlo Method}}
\author{Jian-Sheng Wang$^1$ and Robert H. Swendsen$^2$ \\
\\
$^1$Department of Computational Science,\\
National University of Singapore, Singapore 119260,\\
Republic of Singapore\\
and\\
$^2$Department of Physics, Carnegie Mellon University,\\
Pittsburgh, PA 15213, USA}
\date{21 April 2001}
\maketitle

\begin{abstract}
We present a formalism of the transition matrix Monte Carlo method. A
stochastic matrix in the space of energy can be estimated from Monte
Carlo simulation. This matrix is used to compute the density of
states, as well as to construct multi-canonical and equal-hit
algorithms. We discuss the performance of the methods. The results are
compared with single histogram method, multi-canonical method, and
other methods. In many aspects, the present method is an improvement
over the previous methods.

\vskip\baselineskip PACS numbers: 02.70.Tt, 05.10.Ln, 05.50.+q.

Keywords: Monte Carlo method, flat histogram, multi-canonical ensemble.
\end{abstract}

\section{Introduction}

The Monte Carlo technique \cite{binder} has served us well in the
study of equilibrium statistical mechanics and other fields. The
traditional local Monte Carlo method is simple, extremely general, and
versatile. However, there are some intrinsic drawbacks. First, the
convergence of the results to the exact values is slow. The basic
probabilistic nature has limited the Monte Carlo error to decrease as
$1/\sqrt{t}$, where $t$ is Monte Carlo steps or computer time. With
exception of quasi Monte Carlo \cite{qMC} for numerical integration,
as long as we use a probabilistic approach, it does not appear
possible to overcome this barrier. Most of the work to improve the
efficiency of Monte Carlo method has been via variance reduction
\cite{variance-reduction,assaraf}, which reduces the value of coefficient 
in front of the $1/\sqrt{t}$ law. Next, while the traditional Monte
Carlo method is good for computing expectation values such as the
internal energy and its derivatives, it is more difficult to compute
the free energy or entropy \cite{meirovich}.

Over the last few decades, a number of methods have been developed to
compute the density of states. The histogram method
\cite{ferrenberg-swendsen} and the multiple histogram method 
\cite{ferrenberg-swendsen-m} can be regarded from this point of view. 
The multi-canonical method \cite{berg} in some sense is also a
computation of the density of states. Both of these methods involve
the re-weighting of probabilities to construct the canonical
distribution. Oliveira \textsl{et al\/} \cite{oliveira-brasil1}
proposed a broad histogram method, in which the density of states is
also computed from simulation.

If the density of states can be computed with sufficient accuracy,
then most thermodynamic quantities can be obtained with little further
effort. This includes the moments of energy, entropy, and free
energy. Moreover, the results are obtained as a continuous function of
temperature from a single simulation. In this paper, we present such a
method and study its efficiency. This method includes the use of a
transition matrix \cite{wang-tay-swendsen-PRL}, a stochastic matrix
defined in the space of energy, and a class of related simulation
algorithms \cite{wang-europhys-B,wang-lee,swendsen-int}. The present
method has the elements of both the broad histogram method and
multi-canonical method. The flat-histogram algorithm offers an
effective way to compute density of states $n(E)$ for all energy
$E$. With its multi-canonical element, it also offers fast dynamics
for systems at first-order phase transitions. The use of transition
matrix improves the efficiency of data analysis.

In the next section, we shall discuss the formalism and the essential
aspects of the method. We also present the results of some numerical
tests and discuss the connections of our method with previous
methods. In the Appendices, we give details of a transition matrix
dynamics.

\section{Formalism}

\subsection{Markov Chain Monte Carlo}

The Monte Carlo method aims at generating samples $\sigma $ with
probability distribution $P(\sigma )$, where $\sigma $ is a particular
state of the system.  In the Ising model, which we shall use as a
concrete example, $\sigma $ is a vector of all the spins $\{\sigma
_{1},\sigma _{2},\cdots ,\sigma _{N}\}$, where $\sigma _{i}=\pm 1$. In
the usual application of the Monte Carlo method, the invariant
distribution $P(\sigma )$ is given by the canonical distribution
(Gibbs distribution) $\exp \bigl(-E(\sigma )/k_{B}T\bigr)$. However,
this need not be the case. In the equal-hit ensemble that we shall
discuss later, $P(\sigma )$ is not known, and is not even unique.
Nevertheless, it is still a valid Monte Carlo algorithm that can have
significant advantages.

A sequence of states or samples is generated by a Markov chain
\cite{markov-chain} with transitions between states described by a matrix
$W(\sigma \rightarrow \sigma ^{\prime })$. This is the conditional
probability that state moves to $\sigma ^{\prime }$ given that the
current state is $\sigma $. This matrix is known as a stochastic
matrix and it must satisfy
\begin{equation}
\sum_{\sigma ^{\prime }}W(\sigma \rightarrow \sigma ^{\prime })=1,\qquad
W(\sigma \rightarrow \sigma ^{\prime })\geq 0.  \label{eq-norm}
\end{equation}%
There is considerable freedom in choosing the matrix $W$, but the most
important condition (or criterion) is detailed balance
\begin{equation}
P(\sigma )W(\sigma \rightarrow \sigma ^{\prime })=P(\sigma ^{\prime
})W(\sigma ^{\prime }\rightarrow \sigma ).  \label{eq-detailed-balance}
\end{equation}%
A Markov chain that satisfies the above condition is called a
reversible Markov chain. This condition guarantees the invariance of
the probability $P(\sigma )$ with respect to the transition matrix
$W$, i.e.,
\begin{equation}
\sum_{\sigma }P(\sigma )W(\sigma \rightarrow \sigma ^{\prime })=P(\sigma
^{\prime }).
\end{equation}%
Repeated application of $W$ to an arbitrary probability distribution
makes the resulting probability distribution converge to a fixed
point, $P(\sigma ) $. We shall not elaborate on the condition that an
invariant probability distribution exists and is unique.  Roughly
speaking, we must be able to make transitions in a finite numbers of
steps from any initial state to any final state. This is known as
ergodicity.

The standard Metropolis algorithm \cite{metropolis} is to take
\begin{equation}
W(\sigma \rightarrow \sigma ^{\prime })=S(\sigma \rightarrow \sigma ^{\prime
})\min \left( 1,{\frac{P(\sigma ^{\prime })}{P(\sigma )}}\right) ,\quad
\sigma \neq \sigma ^{\prime },  \label{eq-metroplis-rate}
\end{equation}%
where $S(\sigma \rightarrow \sigma ^{\prime })$ is a selection
function -- a conditional probability of attempting to go to state
$\sigma ^{\prime }$ given that the current state is $\sigma $. Within
the above formulation, it is required that the $S$ matrix is
symmetric,
\begin{equation}
S(\sigma \rightarrow \sigma ^{\prime })=S(\sigma ^{\prime }\rightarrow
\sigma ),
\end{equation}%
although this condition can also be relaxed \cite{hasting}.  Following
Oliveira \cite{oliveira-europhys-b-2}, we call this condition
microscopic reversibility.  The diagonal elements of $W$ are fixed by
the normalization condition, Eq.~(\ref{eq-norm}). Note that the
diagonal elements are not needed explicitly in a computer simulation.

In a computer implementation, a move is selected according to $S$ (e.g.,
pick a site to flip a spin). The move is made if a random number $\xi $
between 0 and 1 is less than the flip rate $\min \bigl(1,P(\sigma ^{\prime
})/P(\sigma )\bigr)$; otherwise, it is rejected, and the original
configuration $\sigma $ is counted once more as the next configuration in a
Monte Carlo move.

Clearly the above formalism is very general. Although the procedure
can be used to sample any distribution, it has its limitations. One
drawback of standard algorithm is that the configurations generated
are correlated.  These correlations severely limit the efficiency of
the method near phase transitions or for models with competing
interactions and many local minima.  A number of methods have been
proposed to address this problem, such as the cluster algorithms
\cite{swendsen-wang-PRL}, the multi-canonical methods
\cite{berg,lee,besold}, replica Monte Carlo \cite{replica}, and simulated
tempering \cite{tempering}. The flat histogram method presented in
this paper is similar in some aspects to the multi-canonical
method. The implementations of flat histogram method and transition
matrix based methods are very simple and efficient.

\subsection{Histogram}

The concept of an energy histogram is essential to all of these
methods.  Other types of histogram of macroscopic quantities can be
easily defined in analogy to the energy histogram and may be useful in
some contexts, such as the joint histogram of energy and total
magnetization. We define the energy histogram (in the case of a
discrete energy spectrum) as the number of instances of each value of
the energy $E$ generated during a Monte Carlo simulation; we denote
the histogram by $H(E)$.

The histogram is important because of its direct relationship to the
probability distribution of the energy in the system being simulated.
If the probability of a state $\sigma $ is given by $P(\sigma )$ for a
given simulation, then
\begin{equation}
h(E)=\sum_{\sigma} \delta_{E(\sigma),E} P(\sigma) =
\sum_{E(\sigma )=E}P(\sigma )
\end{equation}%
is the probability that the system has an energy $E$. If the Monte
Carlo simulation in question generates $m$ configurations, then the
expectation value of the histogram (average of the histogram over an
infinite number of similar Monte Carlo runs) is given by
\begin{equation}
\left\langle H(E)\right\rangle =m\,h(E).
\end{equation}
For the canonical distribution, we have
\begin{equation}
P(\sigma )=f\Bigl(E(\sigma )\Bigr) /Z= Z^{-1} \exp 
\Bigl(-E(\sigma)/k_{B}T\Bigr),
\end{equation}
where $Z$ is the partition function,
\begin{equation}
Z=\sum_{\sigma }\exp \left( -{\frac{E(\sigma )}{k_{B}T}}\right) ,
\end{equation}%
then $h(E)=n(E)f(E)/Z$. We define the density of states (for systems
with discrete energy spectrum)
\begin{equation}
n(E)=\sum_{E(\sigma )=E}1
\end{equation}%
as the number of states with energy $E$.

Note that the configuration dependence of the probability is only
through energy implicitly. Thus, two configurations with the same
energy will have the same probability. We shall call this the
microcanonical property. The transition matrix Monte Carlo to be
discussed below relies on this property crucially, while allowing the
function $f(E)$ to be arbitrary.

The histogram $H(E)$ sampled during a Monte Carlo run (the number of
visits to energy $E$) is an estimator to $h(E)$, i.e., $H(E)\propto
h(E)$. The usual canonical Monte Carlo method is equivalent to using
the number of visits $H(E)$ to compute the moments of $E$ at the
simulation temperature $T_{0}$. The histogram method of Ferrenberg and
Swendsen \cite{ferrenberg-swendsen} is based on the simple observation
that density of states can be estimated (up to a proportionality
constant) by $n(E)\propto H(E)/f(E)$. With this information, the
moments of $E$ can be extrapolated for nearby temperatures as well.

Clearly, if we can determine $n(E)$, then most of the energy related
thermodynamic averages can be determined, such as internal energy,
specific heat, free energy, and entropy. The free energy is given then
by
\begin{equation}
F = - k_B T \ln Z = - k_B T \ln \sum_{E} n(E) \exp(-E/k_B T).
\end{equation}
Other quantity of interest can also be computed if a ``histogram'' (as
a function of $E$) of such quantity is also collected,
\begin{equation}
\langle Q \rangle_T = {\frac{ \displaystyle \sum_{E} \langle Q \rangle_E\;
n(E) \exp(-E/k_B T) }{\displaystyle \sum_{E} n(E) \exp(-E/k_B T) }} \approx {%
\frac{ \displaystyle \sum_{E} \langle Q \rangle_E\; H(E) }{\displaystyle %
\sum_{E} H(E) }}.
\end{equation}
The main objective of this paper is to show that we can determine the
density of states $n(E)$ for the whole range of energy with Monte
Carlo sampling efficiently.

\subsection{Detailed balance for histogram}

The transition matrix defined below serves a dual purpose---for the
computation of the density of states and for the construction of flat
histogram algorithms. There are a number of ways to look at the
transition matrix based methods. We shall take the detailed balance
equation, (\ref{eq-detailed-balance}), as a basic starting
point. Consider all initial states $\sigma $ with energy $E$ and all
final states $\sigma ^{\prime }$ with energy $E^{\prime }$. Each pair
of states $\{\sigma ,\sigma ^{\prime }\} $ has a detailed balance
equation. Some of the equations may be the identity $0=0 $ if the
transition by a single-spin flip is not possible. Summing up the
detailed balance equations for all the states $\sigma $ with a fixed energy
$E$ and all the states $\sigma ^{\prime }$ with a fixed energy $E^{\prime }$,
we have
\begin{equation}
\sum_{E(\sigma )=E}\sum_{E(\sigma ^{\prime })=E^{\prime }}\!\!\!P(\sigma
)\,W(\sigma \rightarrow \sigma ^{\prime })=\sum_{E(\sigma )=E}\sum_{E(\sigma
^{\prime })=E^{\prime }}\!\!\!P(\sigma ^{\prime })\,W(\sigma ^{\prime
}\rightarrow \sigma ).
\end{equation}%
Assuming that the configuration probability distribution is a function
of energy only, i.e., $P(\sigma )\propto f\bigl(E(\sigma )\bigr)$, and
defining the transition matrix in energy as
\begin{equation}
T(E\rightarrow E^{\prime })={\frac{1}{n(E)}}\sum_{E(\sigma
)=E}\sum_{E(\sigma ^{\prime })=E^{\prime }}W(\sigma \rightarrow \sigma
^{\prime }), \label{eq-T-matrix-definition}
\end{equation}%
we have 
\begin{equation}
n(E)f(E)\,T(E\rightarrow E^{\prime })=n(E^{\prime })f(E^{\prime
})\,T(E^{\prime }\rightarrow E).  \label{eq-energy-detailed-balance}
\end{equation}%
As a consequence of $W$ being a stochastic matrix, $T(E\rightarrow
E^{\prime })$ is also a stochastic matrix:
\begin{equation}
\sum_{E^{\prime }}T(E\rightarrow E^{\prime })=1,\qquad T(E\rightarrow
E^{\prime })\geq 0,
\end{equation}%
with the histogram $h(E)=n(E)f(E)/Z$ being the invariant distribution:
\begin{equation}
\sum_{E}h(E)\,T(E\rightarrow E^{\prime })=h(E^{\prime }).
\end{equation}
Similar definition to Eq.~(\ref{eq-T-matrix-definition}) was
introduced in ref.~\cite{dill} in a different context.

\subsection{Broad histogram equation}

Because the matrix $T(E\rightarrow E^{\prime })$ is composed of two
factors, only the second of which depends on the specific ensemble
under consideration, it is convenient to refer all calculations to the
``infinite-temperature'' transition matrix,
\begin{equation}
T_{\infty }(E\rightarrow E^{\prime })={\frac{1}{N}}\langle N(\sigma
,E^{\prime }-E)\rangle _{E},
\end{equation}%
where $N$ is the number of spins, or more generally, the number of
allowed moves from a given state.  If we define $\Delta E=E^{\prime
}-E$, we then have
\begin{equation}
\frac{\left\langle N\left( \sigma ,\Delta E\right) \right\rangle_E }{N}={\frac{%
1}{n(E)}}\sum_{E(\sigma )=E}{\frac{N(\sigma ,\Delta E)}{N}}=\sum_{E(\sigma
)=E}\sum_{E(\sigma ^{\prime })=E^{\prime }}{\frac{S(\sigma \rightarrow
\sigma ^{\prime })}{n(E)}}.
\end{equation}%
In a random single-spin-flip dynamics, $S(\sigma \rightarrow \sigma
^{\prime })$ equals $1/N$ if the two configurations $\sigma $ and $\sigma
^{\prime }$ differ by one spin, and zero otherwise. Thus, the second
summation over $\sigma ^{\prime }$ gives the number $N(\sigma ,E^{\prime
}-E)$ of configurations of energy $E^{\prime }$ that can be reached
from $\sigma $ of energy $E$ by a spin flip. The first summation is
over the configurations with energy $E$, i.e., a microcanonical
average of the quantity $N(\sigma ,\Delta E)$.  The constancy of
$S(\sigma \to \sigma')$ for the nonzero matrix elements is important
for this interpretation of $\langle N(\sigma, \Delta E)\rangle_E$.
The quantity $N(\sigma ,\Delta E)$ is central to the current method,
as well as to the broad histogram method \cite{oliveira-broad}.

Within the single-spin-flip dynamics, the matrix $T$ is then given by
\begin{equation}
T(E\rightarrow E^{\prime })=T_{\infty }(E\rightarrow E^{\prime
})\, a(E\rightarrow E^{\prime }),  \label{eq-transition-matrix}
\end{equation}%
where any flip rate $a(E \rightarrow E^{\prime})$ can be inserted once
$T_{\infty }(E\rightarrow E^{\prime })$ has been determined.
Substituting Eq.~(\ref{eq-transition-matrix}) into the energy detailed
balance equation (\ref{eq-energy-detailed-balance}), we can cancel
$f(E)$ and $a(E\rightarrow E^{\prime })$ as for a valid dynamics which
generates distribution $P(\sigma )=f\bigl(E(\sigma )\bigr)/Z$, we have the usual
detailed balance,
\begin{equation}
{\frac{a(E\rightarrow E^{\prime })}{a(E^{\prime }\rightarrow E)}}={\frac{%
f(E^{\prime })}{f(E)}}.
\end{equation}%
The final equation is known as broad histogram equation, initially
presented by Oliveira \textsl{et al\/} \cite{oliveira-brasil1},
\begin{equation}
n(E)\langle N(\sigma ,E^{\prime }-E)\rangle _{E}=n(E^{\prime })\langle
N(\sigma ^{\prime },E-E^{\prime })\rangle _{E^{\prime }}.
\label{eq-broad-histogram}
\end{equation}%
In terms of the transition matrix notation, this becomes
\begin{equation}
n(E)T_{\infty }(E\rightarrow E^{\prime })=n(E^{\prime })T_{\infty
}(E^{\prime }\rightarrow E).  \label{Tmatrix-broad-hist}
\end{equation}%
The name ``broad histogram'' equation is historical and clearly a
misnomer.  The above equation has a very simple
interpretation. Consider all pairs of states $\sigma $ with energy $E$
and states $\sigma ^{\prime }$ with energy $ E^{\prime }$ such that
the moves (or transitions) between $\sigma $ and $\sigma ^{\prime }$
are allowed. These states correspond to states for which the matrix
elements in $S$ are non-zero. Due to the microscopic reversibility, if
$\sigma $ to $\sigma ^{\prime }$ is allowed, so is the reverse move
from $\sigma ^{\prime }$ to $\sigma $. There are two ways to count the
total number of moves, summing up from states with energy $E$ or
summing up from states with energy $E^{\prime }$.  The state $\sigma $
has $N(\sigma ,E^{\prime }-E)$ ways to move into energy $E^{\prime
}$. The total number of moves to energy $E^{\prime }$ from all states
with energy $E$ is $\sum_{E(\sigma )=E}N(\sigma ,E^{\prime }-E)$. By
the reversibility requirement, we must have
\begin{equation}
\sum_{E(\sigma )=E}N(\sigma ,E^{\prime }-E)=\sum_{E(\sigma ^{\prime
})=E^{\prime }}N(\sigma ^{\prime },E-E^{\prime }).  \label{eq-N-eq}
\end{equation}%
A microcanonical average of any quantity is defined by 
\begin{equation}
\langle Q\rangle _{E}={\frac{\displaystyle\sum_{E(\sigma )=E}Q(\sigma )}{%
\displaystyle\sum_{E(\sigma )=E}1}}={\frac{1}{n(E)}}\sum_{E(\sigma
)=E}Q(\sigma ).
\end{equation}%
Using this definition, the previous equation (\ref{eq-N-eq}) is
reduced to Eq.~(\ref{eq-broad-histogram}). This argument is first put
forth by Oliveira \cite{oliveira-europhys-b-2} and by Berg and
Hansmann \cite{berg-hansmann-europhys-b}. Clearly, the result does not
depend on what type of moves we use, so long as it satisfies the
reversibility condition.

\subsection{$\mathbf{TTT}$ identity}

The detailed balance condition imposes a restriction on the transition
matrix, which we call the $TTT$ identity. Consider three distinct
energy levels, $E$, $E^{\prime }$, and $E^{\prime \prime }$, for which
energy transition matrix elements among them are nonzero. Let us write
down three equations associated with the transitions among them:
\begin{eqnarray}
h(E)\,T(E\rightarrow E^{\prime }) &=&h(E^{\prime })\,T(E^{\prime
}\rightarrow E), \\
h(E^{\prime })\,T(E^{\prime }\rightarrow E^{\prime \prime }) &=&h(E^{\prime
\prime })\,T(E^{\prime \prime }\rightarrow E^{\prime }), \\
h(E^{\prime \prime })\,T(E^{\prime \prime }\rightarrow E)
&=&h(E)\,T(E\rightarrow E^{\prime \prime }).
\end{eqnarray}%
Multiplying the left and right sides of the three equations together
and assuming that the product $h(E)h(E^{\prime })h(E^{\prime \prime
})$ is nonzero, we can cancel this factor from the equation and
obtain:
\begin{equation}
T(E\rightarrow E^{\prime })T(E^{\prime }\rightarrow E^{\prime \prime
})T(E^{\prime \prime }\rightarrow E)=T(E^{\prime }\rightarrow
E)T(E\rightarrow E^{\prime \prime })T(E^{\prime \prime }\rightarrow
E^{\prime }).  \label{eq-TTT-identity}
\end{equation}%
This is the $TTT$ identity. The importance of this equation is that it
does not require the knowledge of the stationary distribution to check
for agreement of the data with the condition of detailed
balance. While for normal Monte Carlo simulation, the detailed balance
is built-in directly to the transition matrix $W$, this is not the
case for some of the transition matrix Monte Carlo schemes that have
been proposed. One implication of detailed balance violation is that
the microcanonical property that all states with the same energy have
the same probability is violated. This detailed balance violation for
the initial choice of Oliveira's broad histogram dynamics, a
particular choice of the transition rate $W$, has been demonstrated
explicitly for small systems \cite{wang-europhys-B}.

The significance of the $TTT$ identity is that given the probability
$h$ of energy having value $E$, if we can predict $h^{\prime\prime}$
at energy $E^{\prime\prime}$ by the detailed balance equations in two
ways, one directly from $E$ to $E^{\prime\prime}$, one by two hops,
from $E$ to $E^{\prime}$, and then $E^{\prime}$ to $E^{\prime\prime}$,
then the $TTT$ identity guarantees that the results are exactly the
same. That is,
\begin{equation}
h^{\prime\prime}= h \: {\frac{ T(E \to E^{\prime\prime}) }{%
T(E^{\prime\prime}\to E) }},
\end{equation}
and 
\begin{equation}
\tilde h^{\prime\prime}= h^{\prime}\: {\frac{ T(E^{\prime}\to
E^{\prime\prime})}{T(E^{\prime\prime}\to E^{\prime})}}, \quad h^{\prime}= h
\:{\frac{ T(E \to E^{\prime}) }{T( E^{\prime}\to E)}}.
\end{equation}
The $TTT$ identity says that the two predictions based on the detailed
balance are equal, $h^{\prime\prime}= \tilde h^{\prime\prime}$.

Detailed balance implies $TTT$ identity. Is the reverse also true?
I.e., given a complete set of $TTT$ identities, do they imply detailed
balance in the sense of equation $h(E)T(E\rightarrow E^{\prime
})=h(E^{\prime })T(E^{\prime }\rightarrow E)$ for all $E$ and
$E^{\prime }$? The answer is yes. The $TTT$ identity turns out to
guarantee a consistent (detailed balance) solution involving three
jumps, say, $E$ to $E^{\prime }$, to $E^{\prime \prime }$, to
$E^{\prime \prime \prime }$, versus $E$ to $E^{\prime \prime \prime }$
directly when such jumps are allowed. Therefore, $TTTT$ identities
that follow from detailed balance are automatically fulfilled and
neither provide further information, nor require separate
proof. Naturally, when we consider adding more complex Monte Carlo
moves, either to improve the efficiency of a calculation or to reflect
the nature of a more complex model, more $TTT$ identities are
generated and identities with more factors of $T$ are automatically
satisfied.

We define a quantity to measure the detailed balance violation for
three energy levels for which the transitions among them are nonzero
as
\begin{equation}
v=\left| 1-{\frac{\hat{T}(E\rightarrow E^{\prime })\hat{T}(E^{\prime
}\rightarrow E^{\prime \prime })\hat{T}(E^{\prime \prime }\rightarrow E)}{%
\hat{T}(E^{\prime }\rightarrow E)\hat{T}(E\rightarrow E^{\prime \prime })%
\hat{T}(E^{\prime \prime }\rightarrow E^{\prime })}}\right| ,
\end{equation}%
where $\hat{T}(\cdots )$ is Monte Carlo estimate of $T(\cdots )$. For
a single-spin-flip dynamics with the Metropolis rate, the energy
transition matrix is given by Eq.~(\ref{eq-transition-matrix}), with
$a(E\rightarrow E^{\prime })=\min \bigl(1,f(E^{\prime })/f(E)\bigr)$,
thus the above equation is equivalent to
\begin{equation}
v=\left| 1-{\frac{\langle N(\sigma ,E^{\prime }-E)\rangle
_{E}\langle N(\sigma ^{\prime },E^{\prime \prime }-E^{\prime })\rangle
_{E^{\prime }}\langle N(\sigma ^{\prime \prime },E-E^{\prime \prime
})\rangle _{E^{\prime \prime }}}{\langle N(\sigma ^{\prime
},E-E^{\prime })\rangle _{E^{\prime }}\langle N(\sigma ,E^{\prime
\prime }-E)\rangle _{E}\langle N(\sigma ^{\prime \prime },E^{\prime
}-E^{\prime \prime })\rangle _{E^{\prime \prime }}}}\right| .
\label{eq-detailed-balance-violation}
\end{equation}%
For Ising model where energies are equally spaced, we consider three
levels at $E$, $E+4J$ and $E+8J$, where $J$ is the coupling
constant. A plot of $v$ for Ising model is presented in
ref.~\cite{wang-lee}.

\subsection{Flat histogram dynamics}

If a configuration has probability $P(\sigma) =
f\bigl(E(\sigma)\bigr)/Z$, then the histogram is $H(E) \propto n(E)
f(E)$. A flat histogram is obtained if we take $f(E) \propto
1/n(E)$. A single spin flip with the flip rate $\min\bigl(1,
n(E)/n(E^{\prime})\bigr)$ can be used to do the simulation.  Lee's
reformulation of multi-canonical method \cite{lee} is essentially
this. The trick there is to determine $n(E)$ efficiently
\cite{berg-JSP}.

From the equation describing detailed balance for the transition
matrix, (\ref{Tmatrix-broad-hist}), we can write the acceptance rate
as
\begin{equation}
a(E\rightarrow E^{\prime })=\min \left( 1,{\frac{T_{\infty }(E^{\prime
}\rightarrow E)}{T_{\infty }(E\rightarrow E^{\prime })}}\right) .
\label{eq-flat-histogram-rate}
\end{equation}%
This is the first equation derived for flat histogram dynamics,
although we will show below that it is not unique. This rate is first
proposed in ref.~\cite{wang-europhys-B}, and is independently
discovered by Li \cite{Li-thesis}. Unlike the quantity $n(E)$, a good
approximation is already available in the very beginning, since we can
use the instantaneous value $N(\sigma, E'-E)$ as a preliminary
estimate for $T_{\infty }(E\rightarrow E')$. A cumulative average of
contributions to $T_{\infty }(E\rightarrow E')$ can be used as a
convenient, and remarkably accurate, approximation for the
microcanonical average. We shall discuss how good this scheme is in a
later section.

There is another equivalent way of looking at the problem. Consider
the energy detailed balance equation in the form
\begin{equation}
h(E)T_{\infty }(E\rightarrow E^{\prime })a(E\rightarrow E^{\prime
})=h(E^{\prime })T_{\infty }(E^{\prime }\rightarrow E)a(E^{\prime
}\rightarrow E).  \label{eq-histogram-detailed-balance}
\end{equation}%
If we require that the histogram is a constant, $h(E)=h(E^{\prime })=
\mathrm{const}$, then the spin-flip rate must satisfy the following equation,
\begin{equation}
\frac{a(E\rightarrow E^{\prime })}{a(E^{\prime }\rightarrow E)}=\frac{%
T_{\infty }(E^{\prime }\rightarrow E)}{T_{\infty }(E\rightarrow
E^{\prime })}.
\end{equation}%
Clearly, Eq.~(\ref{eq-flat-histogram-rate}) satisfies the above
equation.  Moreover, there is a whole family of choices of the
transition rates. Some of the choices are given in
Table~\ref{tb-rates}.

\subsection{$\mathbf{N}$-fold way}

The standard Metropolis algorithm contains two steps. First, a move is
proposed. Next, this move is accepted with probability $a(E\rightarrow
E^{\prime })$ or rejected with probability $1-a(E\rightarrow E^{\prime
})$, where $0\leq a(E\rightarrow E^{\prime })\leq 1$. Is it possible
to always accept a move without sampling the same configuration
repeatedly? The answer is yes, if we are willing to keep extra
information during the simulation.  In the flat histogram method, this
extra information is already there for free. It is precisely $N(\sigma
,\Delta E)$.

In an $N$-fold way simulation (also known as event-driven simulation)
\cite{bortz-kalos-lebowitz}, we do not change the dynamics; it is fully
equivalent to the usual single-spin-flip dynamics. However, there is a
substantial improvement in efficiency in those situations for which
the rejection rate is high. The method begins by computing the total
probability that a move would be accepted with the standard
approach. For the Ising model, this probability is
\begin{equation}
A=A(\sigma )=\sum_{i=1}^{N}{\frac{1}{N}}\,a\left( E(\sigma )\rightarrow E(%
\bar{\sigma}^{i})\right) ,  \label{eq-A}
\end{equation}%
where the factor $1/N$ is due to the fact that each spin located at
site $i$ is picked up with probability $1/N$.  The notation
$\bar{\sigma}^{i}$ refers to a configuration with the spin at site $i$
reversed in sign. The quantity $A$ is the probability that any spin is
flipped. Since the flip rate depends on the initial and final energies
only, we can simplify the above equation as
\begin{equation}
A=\sum_{\Delta E} {N(\sigma, \Delta E) \over N} a(E\rightarrow E+\Delta
E).
\end{equation}%
We divide the possible moves into classes according to their energy
increment $\Delta E$. Within a given class, each spin has the same
flipping probability. We now set the probability of the class with
energy change $\Delta E$ being chosen as
\begin{equation}
P(\Delta E)= { 1 \over A N} N(\sigma, \Delta E) a(E\rightarrow E+\Delta E).
\end{equation}%
A spin in this class is then chosen at random and flipped with
probability one. As a practical consideration in designing algorithms,
we note that the condition that $a(E\rightarrow E^{\prime })$ must be
between 0 and 1 can now be relaxed because of the normalization by
$1/A$ in this equation.

In the original algorithm \cite{bortz-kalos-lebowitz}, Monte Carlo
time is rescaled to make the dynamics equivalent to that of the usual
algorithm. For the purpose of calculating averages, we will reweight
each configuration to achieve the same effect. Each configuration
$\sigma $ in the original single spin flip has equal weight. The Monte
Carlo average of a quantity $Q$ is computed as
\begin{equation}
\langle Q\rangle ={\frac{1}{m}}\sum_{t=1}^{m}Q_{t}.  \label{eq-average}
\end{equation}%
Here $Q_{t}$ at step $t$ and the subsequent steps could be the same,
due to the possibility of rejecting a move. The probability that a
move is rejected is $R=1-A$. Thus the life-time of a configuration has
the probability distribution
\begin{equation}
P_{t}=(1-R)R^{t-1},\quad t=1,2,3,\cdots .
\end{equation}%
The average life-time for configuration $\sigma $ is then 
\begin{equation}
\bar{t}=\sum_{t=1}^{\infty }t\,P_{t}={\frac{1}{1-R}}={\frac{1}{A}}.
\end{equation}%
In the $N$-fold-way simulation, each step generates a distinct state
from the immediate preceding one; each of these states is supposed to
last for a during of $\bar{t}$, on average. Thus in replacing
Eq.~(\ref{eq-average}) the correct formula for statistical average
with $N$-fold-way simulation is
\begin{equation}
\langle Q\rangle =\sum_{t=1}^{m}{\frac{Q_{t}}{A}}\bigg/\sum_{t=1}^{m}{\frac{1%
}{A}}.  \label{eq-N-fold-average}
\end{equation}

Naively the computation of $N(\sigma, \Delta E)$ seems to require
$O(N)$ basic steps at each single-spin-flip move. But the effect of
changing configuration by a flip is local, involving only the site in
question and its neighboring sites. We only need to compute the
changes in $N(\sigma,\Delta E)$. Thus each move takes $O(1)$ in
computer time. It is few times slower than a corresponding
straightforward single-spin-flip program. The $N $-fold way does
require extra memory, as a list of spins for each class is required in
order to be able to pick a spin from the class with a computer time of
$O(1)$.

\subsection{Equal-hit algorithms}

The flat histogram ensemble in some sense is the best ensemble to
evaluate the density of states, for each energy level is sampled with
the same frequency. However, as we have seen in the $N$-fold way, this
is not entirely true, as some configurations are weighted more than
others. The equal-hit algorithms \cite{swendsen-int} generate fresh
configurations with equal probability at each energy. There is a very
interesting aspect of these algorithms---the histograms in such
algorithms are not unique and depend on the details of the dynamics.

The energy histogram in the normal single-spin-flip dynamics (that is,
not in the $N$-fold way) is computed as
\begin{equation}
h(E)=\langle \delta _{E(\sigma ),E}\rangle ,
\end{equation}%
i.e., the contribution to the histogram from the configuration $\sigma
$ of energy $E(\sigma )$ is 1 for $E=E(\sigma )$ and zero for other
energies. The angular brackets refer to Monte Carlo average. In the
$N$-fold way, the statistics are weighted with $1/A$ to get the
equivalent result in the original dynamics, so we have
\begin{equation}
h(E)=\left\langle \delta _{E(\sigma ),E}{\frac{1}{A(\sigma )}}\right\rangle
_{\mathrm{N}}\Biggm/\left\langle {\frac{1}{A(\sigma )}}\right\rangle _{%
\mathrm{N}}.
\end{equation}%
The angular brackets indicate simple arithmetic average over the
samples generated in an $N$-fold way,
c.f. Eq.~(\ref{eq-N-fold-average}). We put a subscript N to emphasize
the fact that the average is over $N$-fold way samples. We define the
hits as
\begin{equation}
u(E)=\langle \delta _{E(\sigma ),E}\rangle _{\mathrm{N}}.
\end{equation}%
This quantity measures the average number of fresh configurations
generated at each energy, since in the $N$-fold way, each
configuration $\sigma $ in the sample is distinct from the previous
sample due to the fact that there is no rejection in $N$-fold way. We
can relate the histogram to the hits by
\begin{equation}
h(E)=u(E)\langle 1/A\rangle _{E,\mathrm{N}}\bigm/\langle 1/A\rangle _{%
\mathrm{N}},  \label{eq-histogram-hit}
\end{equation}%
where 
\begin{equation}
\langle 1/A\rangle _{E,\mathrm{N}}={\frac{\left\langle \delta _{E(\sigma ),E}%
{\frac{1}{A(\sigma )}}\right\rangle _{\mathrm{N}}}{\langle \delta _{E(\sigma
),E}\rangle _{\mathrm{N}}}} = \langle A \rangle_{E}^{-1}
\end{equation}%
is the average of inverse acceptance rate for states of fixed energy
in the $N$-fold way samples. Note that this is not the same as the
microcanonical average of $1/A$, which in general is
\begin{equation}
\langle Q(\sigma )\rangle _{E}={\frac{\langle Q(\sigma )/A\rangle _{E,%
\mathrm{N}}}{\langle 1/A\rangle _{E,\mathrm{N}}}}.
\end{equation}%
Substituting Eq.~(\ref{eq-histogram-hit}) into energy detailed balance
equation, (\ref{eq-histogram-detailed-balance}), we have
\begin{eqnarray}
u(E)\langle 1/A\rangle _{E,\mathrm{N}}T_{\infty }(E\rightarrow E^{\prime
}) a(E\rightarrow E^{\prime})  = \qquad\qquad\qquad\nonumber \\
\qquad\qquad\qquad  u(E^{\prime })\langle 1/A
\rangle _{E^{\prime },\mathrm{N}}T_{\infty }(E^{\prime }\rightarrow E)
a(E^{\prime} \rightarrow E). 
\end{eqnarray}%
Equal-hit is realized if the flip rate satisfies the above equation
with $u(E)=u(E^{\prime })=\mathrm{const}$. For example,
\begin{equation}
a(E\rightarrow E^{\prime })=\min \left( 1,{\frac{\langle 1/A\rangle
_{E^{\prime },\mathrm{N}}T_{\infty }(E^{\prime }\rightarrow E)}{\langle
1/A\rangle _{E,\mathrm{N}}T_{\infty }(E\rightarrow E^{\prime })}}\right) .
\label{eq-hit-rate}
\end{equation}%
Other choices are also possible; some of them are given in 
Table~\ref{tb-rates}.

We note that since the transition rate depends on the underlying
dynamics through $A$, there is no guarantee that the histogram $h(E)$
is unique. In fact, in the equal-hit dynamics, $h(E)$ is not known,
and has to be determined self-consistently through the equal-hit
dynamics. We note that $A$ is defined by Eq.(\ref{eq-A}) using $a$,
which uses $A$ in Eq.~(\ref{eq-hit-rate}). There is also a peculiarity
that $h(E)$ diverges for some choices of the transition rates for the
ground states.

\section{Numerical Tests}

In this section, we evaluate the performance of various proposed
algorithms.  In Table~\ref{tb-rates}, we list a dozen possible choices
of flip rates, $a(E\rightarrow E^{\prime })$. For each rate, the
simulation can be implemented with or without the $N$-fold way.

As the flip rates require the knowledge of exact microcanonical
average, which is not available, we use the crucial approximation
\begin{equation}
T_{\infty }(E\rightarrow E+\Delta E)\approx {\frac{1}{H(E)N}}%
\sum_{t=1}^{m}\delta _{E(\sigma ^{t}),E}N(\sigma ^{t},\Delta E),
\end{equation}%
where energy histogram $H(E)$ is the number of samples generated for a
given energy, $H(E)=\sum_{t=1}^{m}\delta _{E(\sigma ^{t}),E}$, $m$ is
the total number of samples generated so far, and $\sigma^{t}$ is the
configuration at step $t$ in the algorithm.  We collect a sample after
every attempt of moves. The above expression is suitable for the
normal single-spin-flip algorithm. In $N$-fold way, the statistics
have to be weighted by $1/A(\sigma )$. Whenever information is not
available, we set the flip rate to 1. This biases the system to visit
unexplored energy levels.

In the above method, the simulation is started automatically, without
an iteration process. This bootstrap is efficient and is also correct
in the sense that it converges. We shall call the above method
iteration 0.  Strictly speaking, iteration 0 need not be a valid Monte
Carlo algorithm as the transition rates are fluctuating quantities,
thus the normal Markov chain theory for convergence does not
apply. However, numerical results do support convergence, although a
rigorous proof is lacking.

\subsection{Histograms}

\begin{figure}[t]
\epsfxsize=0.8\textwidth\epsfbox{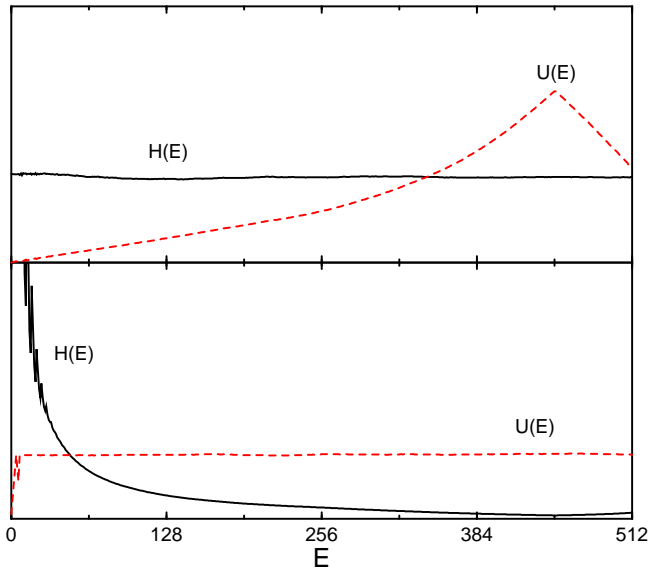}
\caption[histogram]{Energy histogram $H(E)$ and number of hits $U(E)$ in 
$N$-fold-way algorithms applied to the 8-state ferromagnetic Potts
model on a $16 \times 16$ lattice with $10^7$ Monte Carlo steps. The
top part is the standard flat histogram algorithm 0; the bottom part
is the equal-hit algorithm 1. The vertical scales are arbitrary.}
\label{fig-histo}
\end{figure}

Figure~\ref{fig-histo} presents energy and hit histograms for a $16
\times 16 $ square lattice for an 8-state ferromagnetic Potts model.
As expected, the energy histogram is flat for the flat histogram
algorithm. The hits are roughly proportional to the number of
different configurations generated. This is not exactly true since in
the $N$-fold way, we can only guarantee that the next configuration is
different from the immediate preceding one. The new configuration can
be the same with configurations in earlier steps. By requiring that
the hits are equal for all energies, we obtain the equal-hit algorithm
with the corresponding energy and hit distribution shown in the lower
part of the figure.

\begin{figure}[t]
\epsfxsize=0.90\textwidth\epsfbox{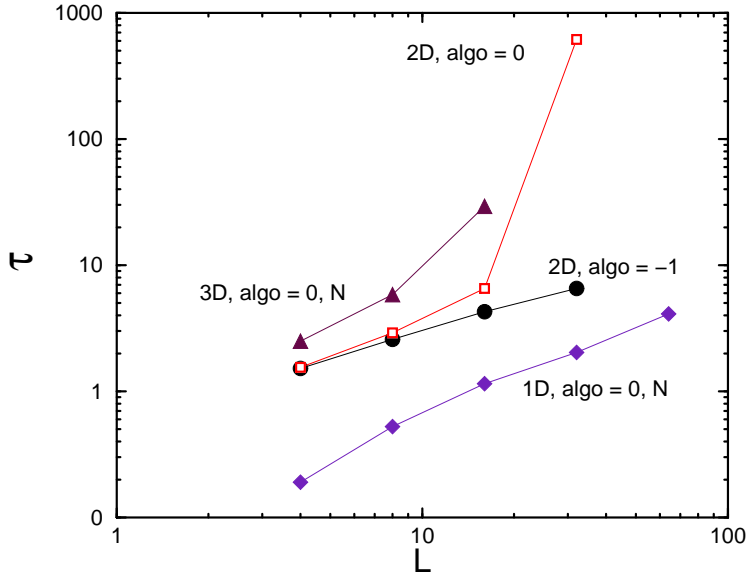}
\caption[flatness of histogram]{Correlation times defined by 
Eq.~(\ref{eq-fluctuation}) for the Ising model versus lattice linear
size $L$ ($=N^{1/d}$). Diamonds are for one-dimensional chain with
algorithm 0 and $N$-fold way; solid circles are for square lattice
with algorithm $-1$; open squares are for square lattice with
algorithm 0; triangles are for cubic lattice with algorithm 0 and
$N$-fold way.}
\label{fig-H-slowing-down}
\end{figure}

Due to the statistical nature of the histogram and also due to the
fact that the energy range is explored similarly to a random walk, the
histogram is not exactly flat, but fluctuates around a mean. During a
simulation of length $t$ of Monte Carlo steps (sweeps), we have
generated $t/\tau$ independent samples, where $\tau$ is correlation
time related to the histogram. These samples are distributed to order
$N$ bins of different energies (for the two-dimensional Ising model,
it is exactly $N-1$ bins). Thus, each bin has about $t/(\tau N)$
independent samples. The relative fluctuation of the histogram is
(asymptotically for large $t$)
\begin{equation}
{\frac{\delta H(E) }{H(E) }} \approx \sqrt{{\frac{ \tau N }{t}} }.
\label{eq-fluctuation}
\end{equation}
Although the above argument applies for the fluctuation between
different runs, it is also reasonable to apply it to the fluctuation
among different energies, since the samples between different energies
are assumed independent. The same analysis also applies to the hits in
equal-hit algorithm.

Equation (\ref{eq-fluctuation}) can be used as a definition for the
correlation time $\tau$. A perfect Poisson process has a constant
correlation time ($\tau = 1$); an ideal random walk in energy is $\tau
\propto N$.  We compute $\tau$ for the nearest neighbor Ising model on
one-dimensional chain, two-dimensional square lattice, and
three-dimensional cubic lattice. We found that numerically
Eq.~(\ref{eq-fluctuation}) is approximately satisfied, with the
correlation time growing with size, $\tau \propto L$ in one dimension,
$\tau \propto L^{0.7}$ in two dimensions, and $\tau \propto L^{1.2}$
or $L^2$ in three dimensions, see Fig.~(\ref{fig-H-slowing-down}). The
algorithm 0 becomes inefficient (slow convergence) for large lattices
in two and three dimensions. We shall discuss this further later on.

\subsection{Rate of convergence}

\begin{table}[t]
\begin{tabular}{|r|c|r|l|}
\hline
No & Rate $a(E \to E^{\prime})$ & $\epsilon_T \sqrt{t}$ & Remark \\ \hline
$-1$ & $\min\Bigl(1, n(E)/n(E^{\prime}) \Bigr)$ & 5.93 & multi-canonical \\ 
0 & $\min\left(1, a/b \right)$ & 6.43 & original flat histogram \\ 
1 & $\min\Bigl(1, a U/(bV) \Bigr)$ & 6.74 & standard equal-hit \\ 
2 & $aU$ & 9.5 &  \\ 
3 & $1/(bV)$ &  & $t^{-1/6}$ slow convergence \\ 
4 & $1/V$ if $a U < b V$, else $a/(bV)$ &  & $t^{-1/3}$ slow convergence \\ 
5 & $a$ & 7.0 &  \\ 
6 & $1/b$ & 7.2 &  \\ 
7 & $a/(a+b)$ & 6.4 &  \\ 
8 & $(a+b)/b$ & 9 &  \\ 
9 & $N/N(\sigma, E^{\prime}-E)$ &  & definitely not converge \\ 
10 & $U/b$ & 8.1 &  \\ 
11 & $a/V$ &  & $t^{-1/5}$ slow convergence \\ 
12 & $U$ if $a U < b V$, else $aU/b$ & 14 &  \\ \hline
\end{tabular}%
\caption[rates]{A list of choices of the flip rates and their errors in
transition matrix with respect to the exact results on a $5 \times 5$
square lattice for the Ising model. In the formula, we have $a =
{\frac{1 }{N}}\langle N(\protect\sigma^{\prime}, E-E^{\prime})
\rangle_{E^{\prime}}$, $b = {\frac{1 }{N}} \langle N(\protect\sigma,
E^{\prime}-E) \rangle_E$, $U =
\langle 1/A(\protect\sigma^{\prime}) \rangle_{E^{\prime},\mathrm{N}}$, $V =
\langle 1/A(\protect\sigma) \rangle_{E,\mathrm{N}}$, where $E$ is the
current energy and $E^{\prime}$ is proposed new energy. In simulation,
the exact microcanonical average $\langle \cdots \rangle_E$ is
replaced by a cumulative average.}
\label{tb-rates}
\end{table}

\begin{figure}[t]
\epsfxsize=0.90\textwidth\epsfbox{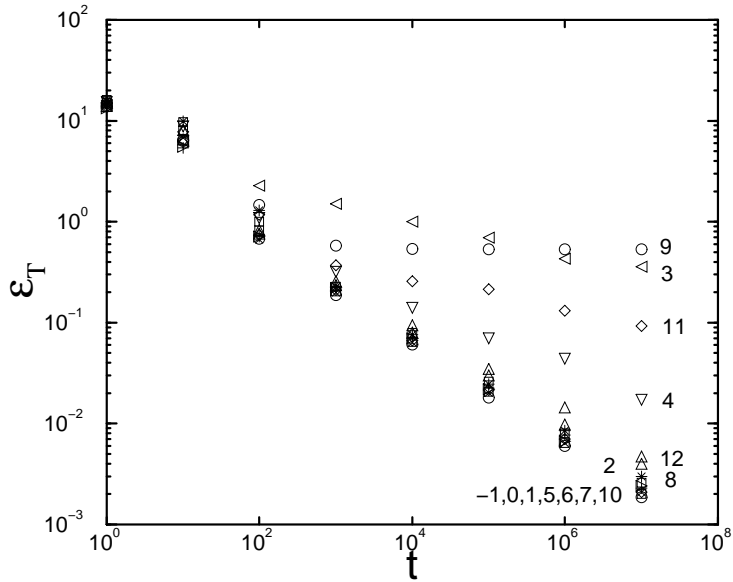}
\caption[tconv]{Average errors in the calculated transition matrix $T_\infty$
on a $5 \times 5$ Ising model in $N$-fold-way simulation. The numbers
correspond to various choices of flip rates given in
Table~\ref{tb-rates}.}
\label{fig-T-converge}
\end{figure}

We tested the zeroth iteration algorithms for convergence to the exact
values for the infinite temperature transition matrix, $T_{\infty
}(E\rightarrow E^{\prime })={\frac{1}{N}}\langle N(\sigma ,E^{\prime
}-E)\rangle _{E}$, on a $5\times 5$ Ising square lattice model. The
exact result $T_\infty$ is obtained by an exhaustive
enumeration. Figure~\ref{fig-T-converge} is a plot of overall error
in the $N$-fold-way simulation data $\hat{T}$, defined by
\begin{equation}
\epsilon _{T}=\sum_{E,E^{\prime }}\left| \hat{T}_{\infty
}(E\rightarrow E^{\prime })-T_{\infty }(E\rightarrow E^{\prime })\right| ,
\end{equation}%
versus Monte Carlo simulation length $t$ (averaged over many
runs). The Monte Carlo times are in units of sweeps ($N$ moves).  The
asymptotic value $\epsilon_{T}\sqrt{t}$ for large $t$ is listed in
Table~\ref{tb-rates}, which characterizes the rate of convergence.

As expected, the best algorithm is algorithm number $-1$, when the
input $n(E)$ is exact. The algorithms (or rates) number 0, 1, and 7
are only slightly worse than the best. Numbers 5 and 6 are the second
best. The rate numbers 3, 4, and 11 may still converge extremely
slowly or not converge at all. We do not understand why this is
so. Number 9 does not converge, which we know since it does not
satisfy detailed balance equation.  Using instantaneous values rather
than average values is clearly wrong.

Errors for single-spin-flip algorithm are about 1.5 larger than with
the $N$-fold-way, or about $1.5^2 \approx 2.3$ less efficient in Monte
Carlo steps. However, since the $N$-fold way is slower in CPU time
than single-spin flips by a factor of 2 or more, the two methods are
comparable in overall efficiency in this particular instance.

From the above results we conclude that even the zeroth iteration
converges to the correct results. This does not mean that the rate of
convergence is uniform in $E$. In fact, we found for large lattices,
the violation of detailed balance, $v$, is large at the two ends of
the energy spectrum \cite{wang-lee}.

\subsection{Determining the density of states}

There are a number of different ways of determining the density of
states.  The matrix $T_{\infty }(E\rightarrow E^{\prime })$ has
eigenvalue 1 with corresponding left eigenvector $n(E)$.  However, the
solutions of the eigenvalue problem are numerically unstable.  The
broad histogram equation, (\ref{eq-broad-histogram}), can be used. In
the simplest application, we can ignore the extra equations, and
consider only these with smallest $\Delta E=E^{\prime }-E$, and obtain
solution from iteration,
\begin{equation}
\ln n(E^{\prime })=\ln n(E)+\ln {\frac{T_{\infty }(E\rightarrow E^{\prime })%
}{T_{\infty }(E^{\prime }\rightarrow E)}}.  \label{eq-simple-iter}
\end{equation}%
Since there are more equations of this type than the unknown $n(E)$,
we can use least-squares method. Our experience suggests that we
should view the problem as an optimization with nonlinear constraints.

There are two possible models for an optimization
solution. Introducing the optimization variable $S(E)=\ln n(E)$,
consider,
\begin{equation}
\mathrm{minimize}\quad \sum_{E,E^{\prime }}{\frac{1}{\sigma _{E,E^{\prime
}}^{2}}}\left( S(E^{\prime })-S(E)-\ln {\frac{T_{\infty }(E\rightarrow
E^{\prime })}{T_{\infty }(E^{\prime }\rightarrow E)}}\right) ^{2},
\label{eq-minimize-n}
\end{equation}%
subject to any known constraints. For the $d$-dimensional Ising model,
we have
\begin{equation}
S(-E)=S(E),\quad S(-dJN)=\ln 2,\quad \sum_{E}\exp \bigl(S(E)\bigr)=2^{N}.
\label{eq-opt-n2}
\end{equation}%
The three conditions are the symmetry between low and high energies,
the degeneracy for the ground states, and the total number of
states. The weight $\sigma ^{2}(E,E^{\prime })$ is the variance of the
Monte Carlo estimates of the quantity $\ln \bigl( T_{\infty
}(E\rightarrow E^{\prime })/T_{\infty }(E^{\prime }\rightarrow E)
\bigr)$. The above minimization problem is essentially a linear
problem. We solve it by an iterative method.

The second, different formulation of the optimization is expressed in
variable $T_{\infty}(E\rightarrow E^{\prime })$,
\begin{equation}
\mathrm{minimize}\quad \sum_{E,E^{\prime }}{\frac{1}{\sigma _{E,E^{\prime
}}^{2}}}\left( T_{\infty }(E\rightarrow E^{\prime })-\hat{T}_{\infty
}(E\rightarrow E^{\prime })\right) ^{2},  \label{eq-opt-t}
\end{equation}%
where $\hat{T}_{\infty }(E\rightarrow E^{\prime })$ is Monte Carlo
estimate with error $\sigma _{E,E^{\prime }}$, and $T_{\infty
}(E\rightarrow E^{\prime })$ is unknown. The minimization is subject
to the conditions
\begin{eqnarray}
0 \leq T_{\infty }(E\rightarrow E^{\prime })\leq 1,&\quad& \sum_{E^{\prime
}}T_{\infty }(E\rightarrow E^{\prime })=1,  \nonumber \\
\prod T_{\infty }(E \rightarrow E^{\prime }) & = & \prod T_{\infty }(E^{\prime
}\rightarrow E).
\end{eqnarray}%
The last one is symbolically a $TTT$ identity, see
Eq.~(\ref{eq-TTT-identity}) for a concrete example. For the Ising
model, there is also an additional symmetry relation
\begin{equation}
T_{\infty }(E\rightarrow E+\Delta E)=T_{\infty }(-E\rightarrow -E-\Delta E).
\end{equation}%
This set of constraints is much more difficult to handle. Intuitively,
this second optimization problem should give better result than the
first one.  However, this is not the case, at least for the
two-dimensional Ising model.  For this model, optimization in
$T_{\infty }(\cdots)$, Eq.~(\ref{eq-opt-t}), gives twice the error
($\epsilon _{n}$ defined below) of the first optimization method. A
simple iteration with Eq.~(\ref{eq-simple-iter}) and $\Delta E=\pm 4J$
has 4 times the error comparing to optimizing solution of
Eq.~(\ref{eq-minimize-n}) and (\ref{eq-opt-n2}).

\begin{figure}[t]
\epsfxsize=0.90\textwidth\epsfbox{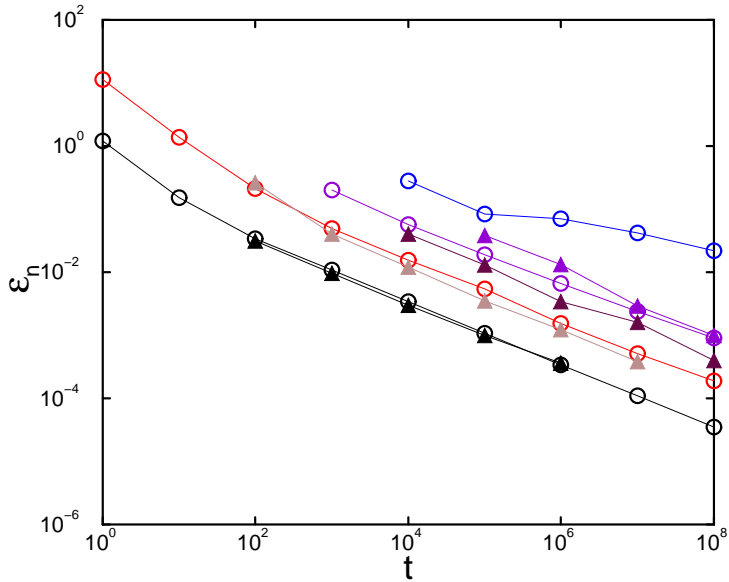}
\caption[n convergence]{Average relative error in density of states 
$\protect\epsilon_n$ as a function of Monte Carlo step $t$. Open
circles are for algorithm 0; solid triangles are for a two-stage
algorithm (algorithm 0 followed by algorithm $-1$). For each set, the
system sizes are $4\times 4$, $8\times 8$, $16\times 16$, and
$32\times 32$, from bottom to top.}
\label{fig-n-conv}
\end{figure}

Figure~\ref{fig-n-conv} is another convergence test plot for algorithm
0 and a two-stage simulation, both with $N$-fold way, on the
two-dimensional Ising square lattices.  In this plot, we consider the
relative error per energy level for density of states,
\begin{equation}
\epsilon_n = {\frac{ 1 }{N-1 }} \sum_E \left| {\frac{ \hat n(E) }{n(E) }} -
1 \right| \approx \langle \bigl| \hat S(E) - S(E) \bigr| \rangle.
\label{eq-n-err}
\end{equation}
The normalization by $N-1$ is to exclude ground state energy and its
symmetric state energy, for which the exact values are imposed. The
exact value $n(E)$ is obtained according to ref.~\cite{beale}; $\hat
n(E)$ is Monte Carlo estimate obtained from solving
Eq.~(\ref{eq-minimize-n}) with $\sigma_{E,E'} = 1$.

We see signs of slower convergence for large lattice sizes for
algorithm 0.  A two-stage simulation improves the efficiency to that
of using the exact density of states in flip rates. We first apply the
algorithm 0 using cumulative average in the flip rates; we then apply
algorithm $-1$ (Lee's method of multi-canonical algorithm), using the
density of states obtained in the first step. Both steps use the same
number of Monte Carlo sweeps.  With the two-stage algorithm, the
slowing down is roughly $\tau _{n}\propto L^{1.5}$, using a similar
definition as given in Eq.~(\ref{eq-fluctuation}).

\begin{figure}[t]
\epsfxsize=0.90\textwidth\epsfbox{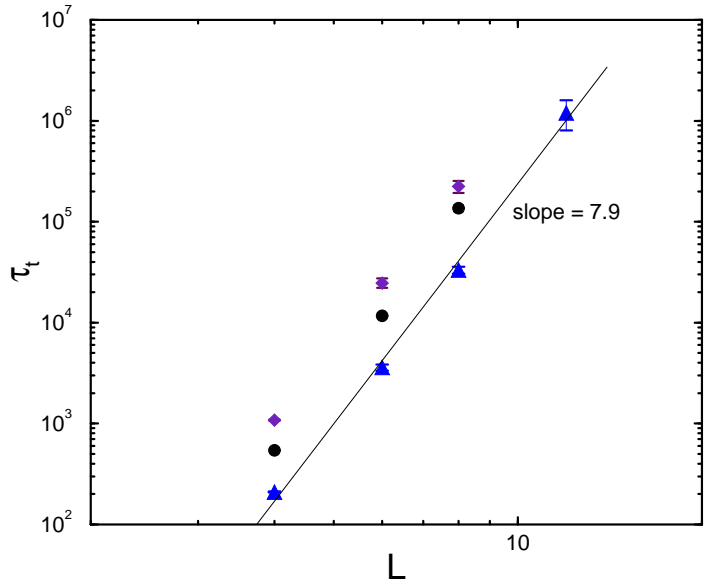}
\caption[tunneling time]{Tunneling time dependences on system sizes for
three-dimensional $\pm J$ spin glasses. The diamonds are from
single-spin flip (without $N$-fold way) with algorithm 0; the circles
are $N$-fold-way algorithm 0; and the triangles are $N$-fold-way algorithm
1.}
\label{fig-tun}
\end{figure}

\subsection{Dynamic characteristics of the algorithms}

We have already presented two correlation times --- $\tau$
characterizes the convergence of the histogram and $\tau_n$
characterizes the convergence of the density of states. Both of them
showed the effect of reduced performance when the size is
increased. By definition of the correlation times $\tau$ and $\tau_n$,
they also measure roughly how many Monte Carlo steps are needed to
generate independent samples. Of course, in $\tau_n$, it also reflects
the effect of data analysis methods.

Another measure of the performance of algorithms is the tunneling
time. In our study, we define the tunneling time as the average Monte
Carlo steps that the system in the lowest energy state goes to the
highest energy state, or vice versa. More precisely, as soon as a
ground state is reached, we record the current Monte Carlo move $t_1$,
and then look for the highest energy, which may happen at $t_2$. We
then look for the ground state again.  The difference $(t_2-t_1)/N$
consists of one sample for the tunneling time.

The tunneling time for the two-dimensional Ising model is very well
described by $\tau_t \approx 0.4 L^{2.8}$ (in units of sweeps). An
ideal random walk in the space of energy would have an exponent of 2
($\tau_t\propto N$ in general). The dynamics is close but not quite
random walk in energy.

The spin glass is one of the most difficult systems to simulate. The
performance of the flat histogram algorithm for the two-dimensional
$\pm J$ spin glass is presented in
ref.~\cite{zhan-lee-wang-physica-A}. In Fig.~\ref{fig-tun}, we show
the tunneling time as a function of system linear size $L$ for three
algorithms (algorithm 0 and 1 with $N$-fold way, and algorithm 0
without $N$-fold way) on the three-dimensional Ising spin glass. The gain
from $N$-fold way as comparing to standard single-spin-flip is by a
constant factor.  The equal-hit algorithm is about a factor of 6
faster in tunneling times than the algorithm 0 without $N$-fold
way. Unfortunately, this gain is not very significant as the
$N$-fold-way program runs few times slower than standard single-spin
flips. The slowing down exponent is about 8, this is comparable to
that of multi-canonical method \cite{berg-hansmann-celik}.

\section{Energy Transition Matrix Dynamics}

The stochastic matrix $W$ describes Monte Carlo dynamics in the space
of spin configurations. Such state space is extremely large,
containing $2^{N}$ states, from which Monte Carlo moves sample only a
very small fraction. On the other hand, we introduced a new stochastic
matrix $T$ in a coarse-grained space of energy. The matrix $W$ and $T$
are related by Eq.~(\ref{eq-T-matrix-definition}). We shall call the
dynamics induced by this stochastic matrix energy transition matrix
dynamics. In discrete time step $i $, the dynamics describes the
evolution of the histogram $h_{i}(E)$ as
\begin{equation}
h_{i+1}(E)=\sum_{E^{\prime }}h_{i}(E^{\prime })T(E^{\prime }\rightarrow E).
\label{eq-discrete-tmmc}
\end{equation}%
What is the significance of this dynamics? The dynamics describes the
change of energy distribution through the following single-spin-flip
moves: given the current state $\sigma $ with energy $E$, pick a new
state $\sigma ^{\prime }$ with the same energy $E$ among all the $n(E)$
degenerate spin states with equal probability, flip a spin according
to the usual spin flip rate as embedded in $W$. As we can see, since
the state changes at random to a completely new state of the same
energy to try another flip, its dynamics is substantially faster than
single spin flip or even cluster flip. Unfortunately, such dynamics is
not realizable on a computer, but it is of interest for comparison
with realizable algorithms.

We can say more about the dynamics given by
Eq.~(\ref{eq-discrete-tmmc}).  Let us first convert the equation into
continuous time which is more convenient for analytic treatment, and
which is a valid description for moderately large system. Introducing
$t = i/N$ and $\Delta t = 1/N$ and define $h(E,i/N) = h_{i}(E)$, we
have
\begin{equation}
h\!\left(E, {\frac{i+1}{N}}\right) - h\!\left(E, {\frac{ i}{N}}\right) =
\sum_{E^{\prime}} h(E^{\prime}, i/N) \Bigl( T(E^{\prime}\to E) -
\delta_{E^{\prime}, E}\Bigr) .
\end{equation}
Dividing both side by $1/N$, taking the limit of large $N$, we have 
\begin{equation}
{\frac{\partial h(E, t) }{\partial t}} = \sum_{E^{\prime}} h(E^{\prime}, t)
\tilde T(E^{\prime}\to E),  \label{eq-t-conti}
\end{equation}
where the continuous time transition matrix $\tilde T$ is related to
the discrete step matrix by $\tilde T(E \to E^{\prime}) = \bigl(T(E
\to E^{\prime}) -\delta_{E, E^{\prime}}\bigr)N$.

Two results were initially reported in
ref.~\cite{wang-tay-swendsen-PRL}.  Detailed derivations will be given
in Appendices. Firstly, an explicit form of $\tilde{T}$ can be given
for the one-dimensional Ising model with Glauber flip rate, as
\begin{eqnarray}
\tilde{T}(k\rightarrow k-1) &=&{\frac{k(2k-1)}{N-1}}(1+\gamma ),
\label{eq-1D-exact-a} \\
\tilde{T}(k\rightarrow k+1) &=&{\frac{(N-2k)(N-2k-1)}{2(N-1)}}(1-\gamma ),
\label{eq-1D-exact-b}
\end{eqnarray}%
where $k=(E/J+N)/4$, $\gamma =\tanh \bigl(2J/(k_{B}T)\bigr)$, and $N$
is the chain length. The diagonal term is computed from the relation
\begin{equation}
\tilde{T}(k\rightarrow k-1)+\tilde{T}(k\rightarrow k)+\tilde{T}(k\rightarrow
k+1)=0,
\end{equation}%
and the rest of the elements $\tilde{T}(k\rightarrow k^{\prime })=0$
if $|k-k^{\prime }|>1$. Secondly, equation (\ref{eq-t-conti}) is
continuous in time but still discrete in energy. We can go one step
further to consider the limit of both time and energy to be
continuous. For transition matrix associated with canonical ensemble,
we found a partial differential equation in such limit as
\begin{equation}
{\frac{\partial h(x^{\prime },t^{\prime })}{\partial t^{\prime }}}={\frac{%
\partial }{\partial x^{\prime }}}\left( {\frac{\partial h(x^{\prime
},t^{\prime })}{\partial x^{\prime }}}+x^{\prime }h(x^{\prime },t^{\prime
})\right) ,  \label{eq-diffusion}
\end{equation}%
where $x^{\prime }$ and $t^{\prime }$ are properly scaled energy
fluctuation and time:
\begin{equation}
x^{\prime }={\frac{E-u_{0}N}{(N\bar{c})^{1/2}}},\quad u_{0}N=\bar{E},
\end{equation}%
and $t^{\prime }=At$ with 
\begin{equation}
A=\lim_{N\rightarrow \infty }\sum_{E}\tilde{T}(E\rightarrow \bar{E}),
\end{equation}%
where $u_{0}$ is the average energy per spin and $\bar{c}=k_{B}T^{2}c$
is the reduced specific heat per spin. This equation,
(\ref{eq-diffusion}), is equivalent to the one-dimensional quantum
harmonic oscillator equation, thus the analytic solutions are readily
obtained.

The most important consequence of this equation is that the relaxation
time is proportional to the specific heat of the system.  This result
can also be seen from a less rigorous point of view.  Since this
artificial dynamics involves a random walk on the probability
distribution of the energy, the characteristic relaxation time will be
proportional to the square of the energy fluctuation, $\left\langle
\left( E-\left\langle E\right\rangle \right) ^{2}\right\rangle $, 
which is in turn proportional to the specific heat.

\section{Connections with Other Methods}

\subsection{Single histogram method}

In the single histogram method \cite{ferrenberg-swendsen}, one
performs a canonical ensemble simulation at a fixed temperature $T_0$,
and collects the histogram $H(E)$. The histogram is proportional to
$n(E) \exp(-E/k_BT_0)$, so an estimate to the density of states is
obtained from
\begin{equation}
n(E) \propto  H(E) \exp\left({\frac{ E }{k_BT_0 }} \right).
\end{equation}
Once we have the density of states, we can use it to evaluate thermodynamic
quantities at any temperature.

\begin{figure}[t]
\epsfxsize=0.90\textwidth\epsfbox{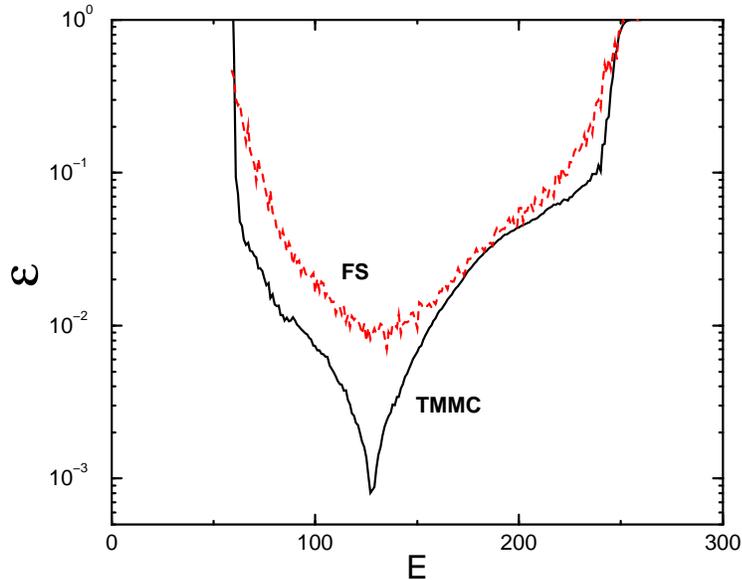}
\caption[TMMC vs single histogram]{Errors in the density of states for a $32
\times 32$ Ising model from a canonical simulation at $k_B T_c/J \approx 2.269$,
with $10^6$ Monte Carlo steps. The dotted line labeled FS is by single
histogram method; the solid line labeled TMMC is obtained by energy
transition matrix with the same simulation. The relative error is
computed from $|\hat n(E)/n(E) - 1|$, where $n(E)$ is the exact value,
$\hat n(E)$ is Monte Carlo estimate, and the error is an average over 48
simulations.}
\label{fig-tmfs-err}
\end{figure}

Unfortunately, since $H(E)$ is approximately a Gaussian function with
mean $\langle E \rangle_{T_0}$ and variance $k_BT_0^2 c N$, where $c$
is the specific heat per spin, the accuracy of the estimates
deteriorates exponentially outside the energy window of order
$\sqrt{N}$, c.f.~Fig.~\ref{fig-tmfs-err}. Detailed error analyses for
energy and specific heat are given in
\cite{ferrenberg-err,newman-err}. The region of good accuracy
coincides with the critical region at a second order phase transition,
so the single histogram method is still an extremely valuable tool to
study phase transitions.

The transition matrix approach can also be used in a way similar to
single histogram method, i.e., collecting the statistics of the
transition matrix in a canonical simulation.  Numerical comparison
suggests that the two methods are comparable in accuracy. In fact, as
we can see in Fig.~\ref{fig-tmfs-err}, for a certain interval, the
transition matrix gives results which are up to 10 times better, but
become comparable or worse outside the limited range of $E$. If we use
the two results to compute the average energy or heat capacity, we
found that the errors are about the same \cite{Li-thesis}. The reason
is that the contributions to errors are dominated by the tails of the
histogram distribution $H(E)$, at these ranges, the density of states
is of comparable accuracy.

It is somewhat disappointing that the single histogram method and
transition matrix Monte Carlo analysis are of the same accuracy. Some
improvement can be made by a careful analysis using Baysian method
\cite{baysian}. But it is unlikely that we can bring about an
improvement of order $\sqrt{N}$ for the accuracy.

\subsection{Multiple histogram and multi-canonical methods}

Both of the multiple histogram method \cite{ferrenberg-swendsen-m} and
the multi-canonical method \cite{berg} give density of states over a
wide range. While the multiple histogram method uses a collection of
standard canonical simulations, the multi-canonical method uses only
one simulation. In reality, a multi-canonical simulation needs to be
iterated few times to converge to the desired ensemble. In this
respect, the flat histogram method takes at most three iterations. The
first iteration already gives excellent results, although there are
noticeable biases for large systems; the second iteration with fixed
flip rates greatly improves the accuracy; the third iteration would
give correct sample average for the transition matrix as well as
correct multi-canonical ensemble.

\begin{figure}[t]
\epsfxsize=0.90\textwidth\epsfbox{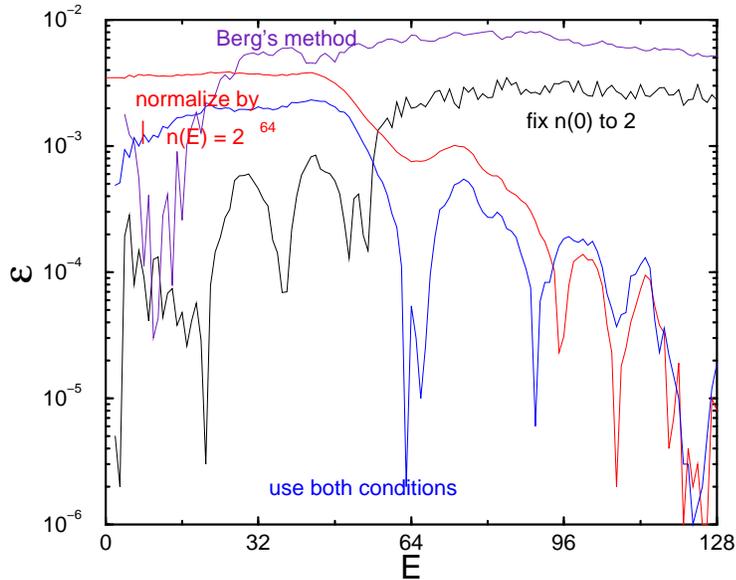}
\caption[TMMC vs multi-canonical]{Errors of a single simulation in the
density of states for a $16 \times 16$ Ising model from algorithm $-1$
(Lee's dynamics). The Monte Carlo steps for the simulation are
$4\times 10^7$. The density of states are calculated by (a) Berg's
method, (b) transition matrix with normalization constraint $\sum_{E}
n(E) = 2^N$, (c) transition matrix with the constraint of groundstate
degeneracy $n(0) = 2$, and (d) transition matrix with both constraints
of (b) and (c). Due to symmetry, only half of the data are plotted.}
\label{fig-2D16lee}
\end{figure}

The additional benefit of using the transition matrix is improved
accuracy comparing to other methods, within the same simulation runs.
Figure~\ref{fig-2D16lee} shows the accuracy of the density of states
for two-dimensional Ising model on a $16\times 16$ lattice. We note
that the accuracy is sensitive to the constraints imposed with the
optimization.  This extra accuracy comes about
due to the nature of the samples that are taken. In histogram or
multi-canonical methods, each new state gives one count in the
histogram, while $N$ counts are collected from each state for the
matrix. Naively, we expect an improvement by a factor of $N$ in terms
of the variance, since each state contributes 1 for the histogram, and
each state contributes a number of $O(N)$ for the transition
matrix. While the accuracy of the transition matrix elements does
improve as the system size increases (may be by $1/\sqrt{N}$ for the
error), as has been pointed out by others
\cite{lima-oliveira-penna-JSP,kastner-promberger-munoz}, this accuracy 
is lost in the density of states. This is due to accumulation of
errors, as the transition matrix elements only determine the ratio of
the density of states, c.f. Eq.~(\ref{eq-simple-iter}). If we use a
simple iteration method starting from the ground state, we see that
this extra accuracy in the matrix elements gets canceled exactly by
the accumulation of errors. However, the optimization methods of
determining $n(E)$ make the error analysis difficult.  Here are some
quantitative comparisons.  We take the exact multi-canonical rate
(algorithm $-1$) in simulation, and collect both the transition matrix
and the histogram.  With transition matrix, the average relative
errors of the density of states defined by Eq.~(\ref{eq-n-err}) for
the two-dimensional Ising model of size $L=4$, 8, 16, 32, and 50 with
$10^6$ Monte Carlo steps in each run are 0.0003, 0.0012, 0.0037,
0.011, 0.024, respectively.  The corresponding results computed by
histograms are 0.0033, 0.010, 0.027, 0.058, 0.10, respectively.  In
general, transition matrix method performs better than histogram based
methods, as we have shown numerically.

\subsection{Transition probability methods}

A proposal to use the transition matrix was given by Smith and Bruce
\cite{smith-bruce} in 1995 in connection with multi-canonical simulation. 
This is further developed by Fitzgerald \textsl{et al\/}
\cite{fitzgerald-picard-silver}. The canonical transition probability (CTP)
method \cite{fitzgerald-picard-silver} also estimates the transition
matrix and uses energy detailed balance equations to estimate the
canonical distribution. In the simplest version, instead of collecting
the histogram $H(E)$, a matrix $H(E\rightarrow E^{\prime })$ is
incremented by 1 for every Monte Carlo move from state with energy $E$
to state with energy $E^{\prime } $. Clearly, this quantity is an
estimator of $h(E)T(E\rightarrow E^{\prime }) $. The transition matrix
is obtained by $H(E\rightarrow E^{\prime })/H(E)$.  Both of the above
methods and the present method are similar in the way that transition
matrix is used.  However, there are two important differences: (1) in
CTP method, only the current move is used for statistics, not all
possible moves of the state; (2) the simulation is performed in
canonical ensemble at a given temperature.

\subsection{Wang-Landau method}

\begin{figure}[t]
\epsfxsize=0.90\textwidth\epsfbox{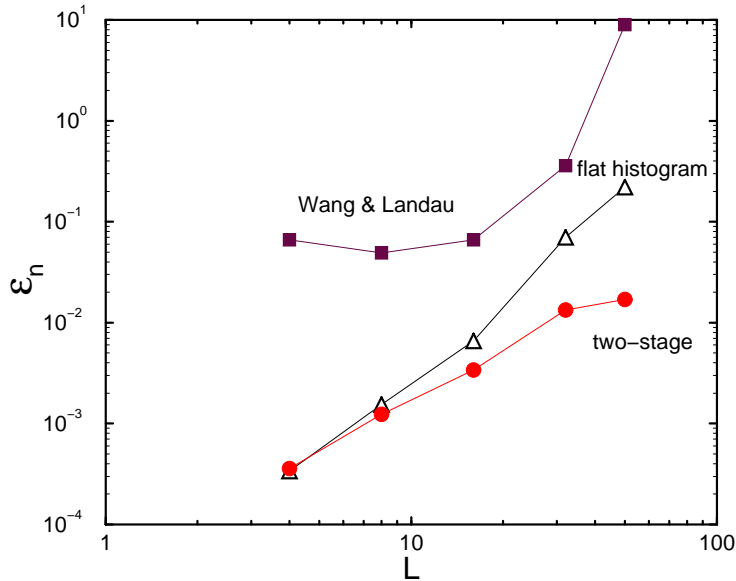}
\caption[Errors in Wang  and Landau]{Average relative error
in the density of states of the two-dimensional Ising model 
for various sizes at fixed Monte Carlo steps.
The two-stage algorithm [circles] used $2\times 10^6$ Monte Carlo
steps ($10^6$ for stage 1 with algorithm 0 and $10^6$ for stage 2 with
algorithm $-1$, using the result of $n(E)$ of stage 1 as input).  The
Wang and Landau method [squares] and the $N$-fold way algorithm 0 flat
histogram [open triangles] used $10^6$ steps, respectively.  The
parameters used in the Wang and Landau's program are 80\% flatness
criterion, $f_0=2.718$, $f_{\min} = 1$, and checking for flatness at
every $10^3$ steps.}
\label{fig-wl-compare}
\end{figure}

Very recently, Wang and Landau \cite{wang-landau} proposed an
intriguing method to determine the density of states.  The dynamics
follows the usual multi-canonical simulation or entropic sampling, by
the single-spin-flip rate $\min\bigl(1, n(E)/n(E')\bigr)$.  However,
$n(E)$ is not a constant, but is updated at each step of trial move
with
\begin{equation}
   n(E) \leftarrow n(E)\, f
\end{equation}
for the current energy $E$.  This is somewhat like the Lee's method of
entropic sampling, but to some extent the updating of the weights are
done at every move.  If $f$ were greater than 1, the algorithm would
never converge, so the idea is to reduce $f$ after some Monte Carlo
steps, by $f \leftarrow f^{1/2}$, for example.  A criterion of
flatness of the histogram was used to determine if $f$ should be
reduced.

Wang and Landau's idea can be adopted in the context of transition
matrix.  For example, we can consider updating the logarithm of
density of states, $S(E) = \ln n(E)$, using the information from the
transition matrix by
\begin{equation}
 S(E) \leftarrow S(E) + \eta \Bigl( S^{\rm pred}(E) - S(E) \Bigr),
\end{equation}
where $0<\eta<1$ is some small parameter and
\begin{equation}
S^{\rm pred}(E) = { 1 \over M } \sum_{E'} \left( S(E') +
  \ln { T_\infty  (E' \to E) \over T_\infty (E \to E') } \right)
\end{equation}
is the predicted logarithmic density of states, based on $M$ possible
hops from $E'$ to the current $E$.  If we already know the ground
state degeneracy, we can fix it to the constant.  Unlike the updating
rule $n(E) \leftarrow n(E) f$ which makes $n(E)$ grow indefinitely,
$S(E)$ will converge to the exact value.  This generalization does
give more accurate results than algorithm 0 if it converges to flat
histogram.  However, it appears to have the problem of sticking to a
Gaussian-like distribution for the histogram for large systems.

We made an extensive test of the accuracy of the random walk algorithm
of Wang and Landau.  In general, the random walk algorithm is far
inferior to the flat histogram algorithm in terms of accuracy and rate
of convergence to flatness, particularly for small systems.  For the
original implementation of the method, once the system passes the
transient period, the error becomes independent of the total Monte
Carlo steps used and the system sizes, and is primarily determined by
how slowly $f$ is reduced.  This feature is useful for its robustness,
particularly for large systems.

In Fig.~\ref{fig-wl-compare} we plot average errors in the density of
states as defined by Eq.~(\ref{eq-n-err}) for fixed Monte Carlo steps
of $10^6$ ($2 \times 10^6$ for the two-stage algorithm).  It is clear
that all methods have bigger errors for larger systems.  But the
random walk algorithm is generally order of magnitudes worse than the
best transition matrix based methods.  The sharp increase of the error
with system sizes from 16 to 32, and to 50 for the random walk
algorithm is an indication that with $10^6$ steps, the system is still
in transients for these sizes.  If $10^7$ Monte Carlo steps are used,
the random walk algorithm comes closer to the accuracy of algorithm 0
at size $L=50$.  The two-stage method (algorithm 0 followed by
algorithm $-1$, both using $N$-fold way) gives the best performance.

\section{Generalization and Some Comments}

The transition matrix approach can be used for continuous degrees of
freedom. In such applications, both the dynamic variables and energy
spectrum have to be discretized. Mu\~noz \textsl{et al\/} have applied
the broad histogram method to the XY model \cite{munoz-XY}. The
important issue here is how to control the numerical error caused by
discretization.

It is straightforward to generalize the transition matrix to more than
one macroscopic variable, such as energy and magnetization; in fact,
this has already been done \cite{lima-multi-parameter} with the broad
histogram method. This approach may have problems, particularly when
the Hamiltonian is complicated. First, the matrix may be too large to
handle in general. Second, with more elements to fill, the statistics
for individual entry is poor. This makes the method less robust.

The transition matrix simulation can be parallelized very efficiently,
where each processor works on separate configurations, using and
updating a common transition matrix.  Each processor can be limited to
work on a range of energies.  The advantage of this is clearly a fast
way of approaching the flat distribution for the histogram.

\section{Conclusion}

Starting from the detailed balance equation, we have formulated the
transition matrix in energy. The infinite-temperature version of this
matrix serves as the basic data from which we determine the density of
states and at the same time is used for construction of flat-histogram
and equal-hit algorithms.  This method of simulation together with
optimization method to determine the density of states offers a better
way of computing thermodynamic quantities by Monte Carlo
simulation. In such an approach, a single simulation produces the
whole function of temperature (or other parameter of the model)
through re-weighting. It is efficient and easy to implement. As the
use of accumulated average for the transition rates causes slow
convergence for large systems, a two-stage iteration is recommended
and is enough to get the best convergence. Dynamically, the
flat-histogram algorithm for long simulations is equivalent to the
multi-canonical method. Using the equal-hit algorithm together with
$N$-fold way offers additional benefits.

\section*{Acknowledgements}

We thank Fugao Wang for providing us his programs for the random walk
simulation.  We also thank Lik Wee Lee, Bernd A. Berg, and Fei Fan for
comments on the manuscript.  J.-S. W. is supported by NUS and
Singapore-MIT Alliance research grants.

\section*{Appendices}

\appendix \setcounter{equation}{0} \renewcommand{\theequation}{A.%
\arabic{equation}}

\section{Exact expression for the energy transition matrix in one dimension}

We consider the single-spin-flip dynamics with a random pick of spins
and the Glauber flip rate
\begin{equation}
a(E \to E + \Delta E) = {\frac{ 1 }{2}} \left[ 1 - \tanh{\frac{ \Delta E }{%
2k_B T }} \right].
\end{equation}
In one dimension for the Ising model, the quantity $\langle N(\sigma,
\Delta E)\rangle_E$ can be evaluated exactly. Let us first define a
set of new variables $n_i = (\sigma_i \sigma_{i+1} + 1)/2$. $n_i$ is 1
for a satisfied bond, and 0 for an unsatisfied bond. The mapping from
$\sigma$ to $n$ is unique modulo an overall spin up-down symmetry. We
assume periodic boundary condition and lattice size $N$ to be
even. There are three possible energy changes, $-4J$, 0, $+4J$. If the
original spin of a site and the spins of two neighboring sites all
have the same sign, it contributes one count for the total $N(\sigma,
4J)$. In terms of $n_i$, it requires two consecutive satisfied bonds,
i.e., $n_i n_{i+1} = 1$. Thus we can express $N(\sigma, +4J) $ in
terms of $n_i$ as
\begin{equation}
N(\sigma, +4J) = \sum_{i=1}^N n_i n_{i+1}.
\end{equation}
Note that only $N-1$ variables $n_i$, $i=1,2, \ldots, N-1$, are
independent (since $\sum_{i=1}^N n_i$ must be even).

A microcanonical average at energy $E$ needs to be carried out. Let us
use $k $ to label the equally spaced energy levels, $k=0, 1, 2,
\ldots, N/2$. Then $E = -NJ + 4Jk$, and $\sum_{i} n_i= n = N-2k$. The
microcanonical average can be expressed as a summation over all $n_i$
subject to $k = (E/J+N)/4$ being an integer constant. Thus we have
\begin{equation}
n(E) \langle N(\sigma, +4J) \rangle_E = 2\mskip-15mu \sum_{\sum_i n_i = n}
\sum_{i=1}^N n_i n_{i+1},
\end{equation}
and similarly 
\begin{equation}
n(E) \langle N(\sigma, -4J) \rangle_E = 2 \mskip-15mu \sum_{\sum_i n_i = n}
\sum_{i=1}^N (1-n_i) (1-n_{i+1}).
\end{equation}
The factor 2 is due to the two-to-one mapping from $\sigma$ to $n$. In
order to compute the above sums, we consider the statistical mechanics
problem of a one-dimensional lattice gas with the Hamiltonian,
\begin{equation}
H = - \epsilon \sum_{i=1}^N n_i n_{i+1} - h \sum_{i=1}^N n_i.
\end{equation}
The partition function of this system (at $1/k_B T=1$) is 
\begin{equation}
Z = \sum_{ \{ n_i \} } \exp\left( \epsilon \sum_{i=1}^N n_i n_{i+1} + h
\sum_{i=1}^N n_i \right).
\end{equation}
Taking the derivative with respect to $\epsilon$, we have 
\begin{equation}
{\frac{\partial Z }{\partial \epsilon }}\Bigg|_{\epsilon = 0} = \sum_{n=0}^N
\left( \sum_{\sum n_i = n} \sum_{i=1}^N n_i n_{i+1} \right) e^{h n} =
\sum_{n=0}^N \Delta^{+}_{{\frac{N-n }{2}}} \mu^n,
\end{equation}
where $\mu = e^h$. The desired quantity is obtained from the
generating function $Z$ as $\langle N(\sigma, +4J) \rangle_E = 2
\Delta^+_{(E/J+N)/4}/n(E)$. The partition function is obtained by the
standard trick of transfer matrix. We find 
\begin{equation}
Z = \lambda^N_{+} + \lambda^N_{-},
\end{equation}
where 
\begin{equation}
\lambda_{\pm} = {\frac{ 1 }{2}} \left( 1 + \mu e^\epsilon \pm \sqrt{ 1 +
4\mu - 2 \mu e^\epsilon + \mu^2 e^{2\epsilon}} \right)
\end{equation}
are the eigenvalues of the matrix 
\begin{equation}
\left[ \matrix{ 1 & \sqrt{\mu} \cr \sqrt{\mu} & e^\epsilon \mu \cr } \right].
\end{equation}
After some algebra, we find 
\begin{equation}
{\frac{\partial Z }{\partial \epsilon }} \Bigg|_{\epsilon = 0} = N \mu^2 ( 1
+ \mu)^{N-2} = N \sum_{n=0}^{N-2} {\frac{ (N-2)! }{n!\, (N-2-n)!}} \mu^{n+2}.
\end{equation}
Thus 
\begin{equation}
n(E) \langle N(\sigma, +4J) \rangle_{(-N+4k)J} = 2 \Delta^+_{k} = {\frac{
2N\, (N-2)! }{(2k)!\, (N-2k-2)! }}.
\end{equation}
A similar derivation from a slightly different Hamiltonian gives
\begin{equation}
n(E) \langle N(\sigma, -4J) \rangle_{(-N+4k)J} = 2 \Delta^-_{k} = {\frac{
2N\, (N-2)! }{(2k-2)!\, (N-2k)! }}.
\end{equation}
Combining the above results with the density of states for the Ising
model, $n(E) = 2\, N!/[(2k)!\, (N-2k)!]$, which is readily obtained by
the combinatorial problem of putting $2k$ unsatisfied bonds in $N$
places, we obtain the expressions given in Eq.~(\ref{eq-1D-exact-a})
and (\ref{eq-1D-exact-b}).

\section{Transition matrix dynamics in the continuum limit}

We start from the dynamical equation
\begin{equation}
{\frac{ \partial\, h(E, t) }{\partial \,t }} = \sum_{E^{\prime}}
h(E^{\prime},t) \tilde T( E^{\prime}\to E),  \label{eq-energy-discrete}
\end{equation}
where time $t$ is continuous but energy $E$ is discrete. The aim is to
consider the continuous energy limit. This limit is natural and is a
very good approximation for large systems. We follow the general
method known as $\Omega$ expansion \cite{vonkampen}. Let us introduce
a new variable,
\begin{equation}
x = {\frac{ E - N u_0 }{\sqrt{N} }},
\end{equation}
where $u_0$ will be determined later. Since $E$ is of order $N$,
naively $x$ is of order $\sqrt{N}$. However, by choosing $u_0$ to be
the average of $E/N$, we cancel the leading $N$ dependence; $x$ is
actually of order 1, measuring the relative fluctuation around
mean. We look for nontrivial solution in variable $x$ in the scaling
limit of $N \to \infty$, keeping $x$ fixed. More precisely, we find
equation in $x$ such that the coefficients of the differential
equation are independent of $N$. Consider the function in terms of $x$
as $\tilde h(x) = h( N u_0 + x \sqrt{N})$. We also write $\tilde T$ in
$x$ as
\begin{equation}
\tilde T_i(x) = \tilde T(N u_0 + x \sqrt{N} + i\,a \to N u_0 + x \sqrt{N}),
\end{equation}
where $i = 0, \pm 1, \pm 2, \cdots, \pm d$ is the change of energy
associated with $E^{\prime}$, and $E^{\prime}= E + i\, a$, $a= 4J$. We
assume a $d$-dimensional Ising model in the derivation. Replacing $E$
by $N u_0 + x \sqrt{N}$, $E^{\prime}$ by $N u_0 + x \sqrt{N} + i\,a$,
Equation~(\ref{eq-energy-discrete}) becomes
\begin{equation}
{\frac{ \partial\, \tilde h(x,t) }{\partial\, t}} = \sum_{i = 0, \pm 1,
\cdots, \pm d} \tilde T_i(x) \tilde h\left(x + {\frac{ i\,a }{\sqrt{N} }}, t
\right).  \label{eq-new-variable}
\end{equation}
The crucial step now is to take Taylor expansion assuming $a/\sqrt{N}$
small, and to find equation that is leading order in the large $N$
limit.  For notational convenient, we shall drop the tildes on $\tilde
T$ and $\tilde h$, which actually denote different functions.

We know in the limit $N \to \infty$, $h(x,t)$ and $T_i(x)$ are smooth
functions in $x$. Then (omitting the variable $t$ for clarity)
\begin{equation}
h\left(x + {\frac{ i\, a}{\sqrt{N}}}\right) = h(x) + {\frac{ i\,a}{\sqrt{N} }%
} h^{\prime}(x) + {\frac{ 1}{2}} \left({\frac{ i\,a}{\sqrt{N} }}\right)^2
h^{\prime\prime}(x) + O(1/N^{3/2}) h^{\prime\prime\prime}(x).
\label{eq-expansion}
\end{equation}
The primes denote derivatives with respect to $x$. We should note that
in the large $N$ limit with $x$ fixed, $h(x)$ and its derivatives do
not contain $N$. The $N$ dependence is made explicit by the above
expansion.  Substituting Eq.~(\ref{eq-expansion}) into
Eq.~(\ref{eq-new-variable}), we have
\begin{equation}
{\frac{\partial h(x) }{\partial t}} = A(x) h(x) + B(x) h^{\prime}(x) + C(x)
h^{\prime\prime}(x) + O(1/\sqrt{N}) h^{\prime\prime\prime}(x),
\end{equation}
where 
\begin{eqnarray}
A(x) & = & \sum_{i} T_i(x), \\
B(x) & = & \sum_{i} T_i(x) {\frac{ i\,a }{\sqrt{N} }}, \\
C(x) & = & {\frac{ 1}{2}} \sum_{i} T_i(x) {\frac{ (i\,a)^2 }{N}}.
\end{eqnarray}
Naively, since $T$ is $O(N)$, we expect $A(x) \sim O(N)$, $B(x)
\sim O(\sqrt{N})$, and $C(x) \sim O(1)$ and the equation does not have a 
well-defined large $N$ limit. But this is not true, due to special
relations among $T_i$.  Two relations are important in the derivation
below to show that both $A(x)$, $B(x)$, and $C(x)$ are of $O(1)$ and
the third derivative can be dropped in the large $N$ limit.

The existence of an equilibrium implies 
\begin{equation}
\sum_{E} \tilde T( E^{\prime}\to E) = 0.
\end{equation}
Expressed in $T_i(x)$, it is 
\begin{equation}
\sum_{i=0, \pm 1, \cdots, \pm d} T_i\left(x - {\frac{ i\,a }{\sqrt{N}}}%
\right) = 0.
\end{equation}
Thus, using $T_0(x)$ from the above 
\begin{eqnarray}
A(x) & = & T_0(x) + \sum_{i \neq 0} T_i(x) \\
& = & \sum_{ i \neq 0 } \left[ T_i(x) - T_i\left(x - {\frac{i\,a }{\sqrt{N}}}
\right) \right] \\
& = & \sum_{i \neq 0} \left[ T_i^{\prime}(x) {\frac{ i\,a }{\sqrt{N} }} - {%
\frac{1}{2}} T_i^{\prime\prime}(x) {\frac{ (i\,a)^2 }{N }} + \cdots \right].
\end{eqnarray}
The last equation above used a Taylor expansion. Since $T_i(x)$ has a
scaling form $T_i(x) \approx N g(x/\sqrt{N})$ in the large $N$ limit,
we find that the $k$-th derivative of $T_i(x)$ at $x=0$ is of order
$T_i^{k}(0) \sim O(N^{1-k/2})$. Thus $T_i^{\prime}(x)$ is of order
$\sqrt{N}$, and we can safely replace $x$ by 0, and find
\begin{equation}
A(x) = \left\{ \sum_i T_i^{\prime}(0) {\frac{ i\,a}{\sqrt{N} }} \right\} +
O\left({\frac{ 1 }{\sqrt{N}}} \right) = A + O\left({\frac{ 1 }{\sqrt{N}}}
\right).
\end{equation}
For $B(x)$, we make an expansion in $x$ for $T_i(x)$ and find 
\begin{eqnarray}
B(x) & = & \sum_{i} T_i(x) {\frac{ i\,a }{\sqrt{N} }} \\
& = & \sum_{i} \left\{ T_i(0) {\frac{ ia }{\sqrt{N} }} + T_i^{\prime}(0) {%
\frac{ i\,a }{\sqrt{N} }} x + {\frac{1}{2}} T_i^{\prime\prime}(0) {\frac{
i\,a }{\sqrt{N} }} x^2 + \cdots \right\},
\end{eqnarray}
where in the last formula, the first term is of order $\sqrt{N}$, the
second term is of order 1, and last term is of order $1/\sqrt{N}$. If
the first term were there, we would have an ill-defined limit. So we
must require that
\begin{equation}
D = \sum_{i} T_i(0) {\frac{i a }{\sqrt{N} }} = 0.
\end{equation}
This is in fact a condition to fix $u_0$. We shall show that this
condition requires $u_0 = \langle E/N \rangle_T$, the canonical
average of energy per spin.

We evoke the energy detailed balance equation,
(\ref{eq-energy-detailed-balance}). In terms of the new variables $x$
and $i$, it is
\begin{equation}
T_i(x) h_{eq}\left(x + {\frac{ia }{\sqrt{N} }} \right) = T_{-i}\left(x + {%
\frac{ia }{\sqrt{N} }} \right) h_{eq}(x).
\end{equation}
Let $\delta = ia/\sqrt{N}$, Taylor expanding the terms involving $\delta$,
we find: 
\begin{equation}
T_i(0) - T_{-i}(0) = - T_i(0) \left[\ln h_{eq}(0) \right]^{\prime}\delta +
T_{-i}^{\prime}(0) \delta + O\left( {\frac{1 }{N}} \right).
\end{equation}
Note that the first term is of order $O(\sqrt{N})$, the second term of
order $O(1)$. It is important to realize that we are looking for the
scaling limit of $N \to \infty$, fixing $x$.

Substituting this equation into the expression for $D$, we find
\begin{equation}
D = \sum_{k=1,2,\cdots,d}\left[ T_k(0) - T_{-k}(0) \right] {\frac{k a }{%
\sqrt{N} }} \approx \sum_{k=1,2,\cdots,d} k^2 T_k(0) [\ln h_{eq}(0)
]^{\prime}{\frac{a^2 }{N}}.
\end{equation}
The requirement that $D=0$ is equivalent to say that $x=0$ is at the
extreme of equilibrium probability distribution.

When the first term in $B(x)$ is set to 0, we have 
\begin{equation}
B(x) = \left\{ \sum_i T_i^{\prime}(0) {\frac{ ia}{\sqrt{N} }} \right\} x +
O\left({\frac{ 1 }{\sqrt{N}}} \right) = A x + O\left({\frac{ 1 }{\sqrt{N}}}
\right).
\end{equation}
The coefficient $C(x)$ is a constant to leading order in $N$: 
\begin{eqnarray}
C(x) & = & {\frac{ 1 }{2}} \sum_{i} T_i(x) {\frac{ (ia)^2 }{N }} \\
& = & {\frac{ 1 }{2}} \sum_i \left\{ T_i(0) {\frac{ (ia)^2 }{N}} +
T^{\prime}(0) {\frac{ (ia)^2 }{N }} x + \cdots \right\} \\
& = & {\frac{ 1}{2}} \sum_i T_i(0) {\frac{ (ia)^2 }{N}} + O( {\frac{1 }{%
\sqrt{N}}}) = C + O( {\frac{ 1 }{\sqrt{N} }}).
\end{eqnarray}

The equilibrium distribution of energy for large system is a Gaussian
distribution with mean $\langle E \rangle_T = N u_0$, and variance
$N\bar c$ where $\bar c$ is reduced specific heat per spin, thus
\begin{equation}
h_{eq}(x) = {\frac{1 }{\sqrt{2 \pi \bar c}}} \exp\left( - {\frac{ x^2 }{%
2\bar c}} \right).
\end{equation}
Substituting this result into the partial differential equation in
equilibrium
\begin{equation}
{\frac{\partial h }{\partial t}} = A h + A x h^{\prime}+ C h^{\prime\prime}=
0,
\end{equation}
we find $C = A \bar c$. This same relation can also be obtained using
detailed balance equation. Changing variables from $t$ to $t^{\prime}=
At$ and from $x$ to $x^{\prime}= x/\sqrt{\bar c}$, we obtain
Eq.~(\ref{eq-diffusion}).

\section{Data file for errors}
The file below is the raw data for various errors.  Formally this is
not part of the paper.  We included here which could be useful for
future benchmarking use.

{%begin small tt font
\fontsize{8pt}{9pt}\selectfont
\begin{verbatim}
Convergence Test Data

MCDIS = -1, SIZE L = 5, N-Fold-Way, 2D Ising, algo -1 to 12
e_T = sum | T - T_ex |

mcs  -1     0      1      2      3      4      5      6      7      8      9      10    11     12  

1    14.961 15.059 15.043 13.995 14.299 15.357 13.330 14.241 14.455 15.883 13.859 16.06 13.776 15.237
10   6.480  8.107  7.795  6.509  6.33   8.83   5.52   5.40   6.79   9.91   6.32   9.68  6.41   8.09
100  0.681  0.819  0.736  0.759  2.28   1.09   0.710  0.713  0.667  1.29   1.47   1.00  1.171  0.825
1000 0.188  0.204  0.210  0.234  1.50   0.320  0.221  0.221  0.199  0.226  0.581  0.227 0.370  0.256
1e4  0.0608 0.0644 0.0663 0.080  1.00   0.14   0.070  0.070  0.064  0.076  0.539  0.079 0.256  0.094
1e5  0.0181 0.0206 0.0215 0.030  0.698  0.070  0.0219 0.0224 0.0201 0.0246 0.5353 0.026 0.2136 0.0349
1e6  0.0060 0.0064 0.0070 0.0098 0.43   0.044  0.0067 0.0077 0.0064 0.0085 0.533  0.008 0.131  0.0144
1e7 0.00186 0.0020 0.0021 0.0040 0.357  0.0172 0.0024 0.0023 0.0021 0.0030 0.5342 0.0026 0.0926 0.0047
1e8         0.00062 0.00066      0.29   0.012                                            0.0842

same as above parameters, but with single-spin-flip (i.e., no N-fold)

mcs  -1     0      1      2      3      4      5      6      7      8      9      10    11     12  
1   16.448  15.26  15.263 15.178 15.307 15.31  15.18  15.31  15.41  15.30  15.31  15.31 15.54  15.30
10   9.329  9.162  9.165  
100  1.286  1.583  1.599
1000 0.287  0.312  0.310
1e4  0.093  0.097  0.095
1e5  0.0287 0.0311 0.0314

           algo no (error at N-fold mcs=1e4)
best        -1     (0.061)
very good   0,1,7  (0.065)
good        5,6,8  (0.07)
OK          2,10   (0.08)
not so good 4,12   (0.1)
bad (converge) 3,11 (>0.25)
don't converge 9   (0.5)


All the rest has MCDIS=0 (no MCS discarded)

2D Ising model, e_n = (1/(N-1)) sum | n(E)/n_ex(E) - 1 | 
                  algo 0 (N-fold-way)
mcs      L=4          L=8           L=16       L=32        L=50
1        1.200       
10       0.152        1.38^
100      0.0340       0.212         1.3^
1000     0.0109       0.0492        0.20       1^
1e4      0.00342      0.0154        0.0568     0.28        0.31 
1e5      0.00108      0.0054        0.0189     0.084       0.23
1e6      0.000338     0.00154       0.0066     0.07        0.22+/-0.02
1e7      0.000110     0.00051       0.0024     0.042       0.191+/-0.0138
1e8      0.000035     0.000190      0.0009     0.022       0.16

^wide distribution, occasionally not converging

tunnel   18.983      144.68         983         6448
time


2D Ising model, e_n = (1/(N-1)) sum | n(E)/n_ex(E) - 1 | 
            algo 1 (N-fold-way, equal-hit)
mcs      L=4          L=8           L=16       L=32        L=50
1        1.20
10       0.142        1.27
100      0.0410       0.178
1000     0.0134       0.051         0.26 
1e4      0.0042       0.0152        0.048      0.18^       0.4
1e5      0.00138      0.0050        0.015      0.04        0.12^
1e6      0.00044
1e7      0.00015

2D Ising model, e_n = (1/(N-1)) sum | n(E)/n_ex(E) - 1 | 
         algo -1 (N-fold-way) (exact rate, best we can do)
mcs      L=4          L=8           L=16       L=32        L=50
1       1.32  
10      0.131       14
100     0.031        0.18 
1000    0.0097       0.038         0.16
1e4     0.0030       0.0125        0.041       0.13        0.23
1e5     0.00096      0.00383       0.0121      0.037       0.07
1e6     0.00031      0.00119       0.0037      0.0114      0.024 
1e7     0.000095     0.00040       0.00122     0.00337     0.007
1e8     0.000032     0.000103      0.00045     0.00091

Ising model, same as above algo -1 N-fold-way, but use histogram to
compute n(E), i.e. Berg/Lee method
mcs      L=4          L=8           L=16       L=32        L=50
1e5     0.010         0.033        0.0733      0.32        0.5
1e6     0.0033        0.0099       0.027       0.058       0.10 

2D Ising model, e_n = (1/(N-1)) sum | ...|
two stage simulation, stage 1, using algo 0, 
                      stage 2, using algo -1 of n(E) generated from stage 0.
same run length for the two stages
mcs       L=4         L=8       L=16           L=32        L=50
100      0.0311       0.26                   
1000     0.0096       0.0404    0.10
10000    0.0030       0.0122    0.0401         0.2         0.5
1e5      0.00100      0.00349   0.0131         0.038       0.075
1e6      0.00036      0.00123   0.0034         0.0133      0.017
1e7                   0.00038   0.0016         0.0029      0.0074
1e8                             0.0004         0.0010

Modified Wang FG/Wang JS method (algo 13'), e_n = (1/(N-1) sum | ...|
using S(E) <- S(E) + eta (S^pred - S(E))
(eta = 0.1)  (two numbers - using Tmatrix/using S(E) directly)
mcs  L=4            L=8            L=16          L=32
100
1000 0.00936/0.358
1e4  0.00306/0.0087 0.39/0.96
1e5  0.00096/0.0028 0.0039/0.015  0.99/0.99
1e6                 0.00115/0.004 0.71/0.99
1e7                 0.00049/0.0014

Wang Fugao's program (fixed MCS, f_max = 2.718, f_min=1, 80%H, check every 1k steps)
e_n = < | n(E)/n_ex(E) - 1 | >
mcs       L=4         L=8       L=16           L=32        L=50
1e4      0.172      10^
1e5      0.065       0.073     1
1e6      0.066       0.049     0.066           0.36        9(+/-4)
1e7      0.055       0.073     0.043           0.05        0.18
1e8      0.069       0.047     0.07            0.025       0.14

Wang FG method e_n (my program algo 13, 80%H, checks 30 times, single-spin-flip)
mcs       L=4         L=8       L=16           L=32        L=50
1e4       0.079       1.62
1e5       0.034       0.071     0.7
1e6       0.012       0.031     0.056         0.42         5.5
1e7       0.0044      0.010     0.025         0.06         0.14
1e8       0.0013                                           0.02
\end{verbatim}
}%end small tt font

\end{document}